\documentclass[]{fairmeta}
\usepackage{textgreek}
\usepackage{float}
\usepackage{booktabs}
\usepackage{tabularx}
\usepackage{makecell}
\usepackage{array}
\usepackage{caption}
\usepackage{siunitx} % for alignment of numbers (optional)
\usepackage{graphicx}
\usepackage{graphicx}
\usepackage{amsmath}
\usepackage[version=4]{mhchem}
\usepackage{booktabs}
\usepackage{caption}

\usepackage[utf8]{inputenc} % allow utf-8 input
\usepackage[T1]{fontenc}    % use 8-bit T1 fonts
\usepackage{hyperref}       % hyperlinks
\usepackage{url}            % simple URL typesetting
\usepackage{booktabs}       % professional-quality tables
\usepackage{amsfonts}       % blackboard math symbols
\usepackage{nicefrac}       % compact symbols for 1/2, etc.
\usepackage{microtype}      % microtypography
\usepackage{xifthen}
\usepackage{multirow}
\usepackage{mciteplus}
\usepackage{epstopdf}
\usepackage{shellesc}

\usepackage{siunitx}
\usepackage{mhchem}
\usepackage{threeparttable}

\newcommand\mcc[1]{\multicolumn{1}{c}{#1}}  % centering a single table cell
\newcommand{\angs}{\text{\AA}}
\newcommand{\op}{OPoly26}

\newcommand{\kindatiny}{\fontsize{6pt}{7.2pt}\selectfont}

\newlength\savewidth\newcommand\shline{\noalign{\global\savewidth\arrayrulewidth
  \global\arrayrulewidth 0.5pt}\hline\noalign{\global\arrayrulewidth\savewidth}}
\newcommand{\tablestyle}[2]{%
    \fontfamily{ptm}\selectfont%
    \let\itold\it%
    \def\it{\itold \fontfamily{ptm}\selectfont}%
    \setlength{\tabcolsep}{#1}\renewcommand{\arraystretch}{#2}\centering\kindatiny%
    \let\citeold\cite%
    \renewcommand{\cite}[1]{\normalfont\fontfamily{ptm}\selectfont\tiny\citeold{##1}}%
}

\newcolumntype{x}[1]{>{\centering\arraybackslash}p{#1pt}}
\newcolumntype{y}[1]{>{\raggedright\arraybackslash}p{#1pt}}
\newcolumntype{z}[1]{>{\raggedleft\arraybackslash}p{#1pt}}
\newcolumntype{w}{>{\centering\arraybackslash}p{18pt}}
\newcolumntype{a}{>{\centering\arraybackslash}p{16pt}}

\definecolor{c0-title-bkg}{HTML}{ffffff}
\definecolor{c0-title-text}{HTML}{000000}
\definecolor{c0-item-bkg}{HTML}{ffffff}
\definecolor{c0-item-text}{HTML}{818589}

\definecolor{c1-title-bkg}{HTML}{d1e2dd}
\definecolor{c1-title-text}{HTML}{005953}
\definecolor{c1-item-bkg}{HTML}{e6efec}
\definecolor{c1-item-text}{HTML}{2d7b6d}

\definecolor{c2-title-bkg}{HTML}{cfe1e1}
\definecolor{c2-title-text}{HTML}{005760}
\definecolor{c2-item-bkg}{HTML}{e4eeed}
\definecolor{c2-item-text}{HTML}{24797b}

\definecolor{c3-title-bkg}{HTML}{cddfe5}
\definecolor{c3-title-text}{HTML}{330704}
\definecolor{c3-item-bkg}{HTML}{facac5}
\definecolor{c3-item-text}{HTML}{330704}

\definecolor{c4-title-bkg}{HTML}{cedce8}
\definecolor{c4-title-text}{HTML}{191b1c}
\definecolor{c4-item-bkg}{HTML}{e2edf4}
\definecolor{c4-item-text}{HTML}{191b1c}

\definecolor{c5-title-bkg}{HTML}{d0d9eb}
\definecolor{c5-title-text}{HTML}{0e1834}
\definecolor{c5-item-bkg}{HTML}{d0e6f9}
\definecolor{c5-item-text}{HTML}{0e1834}

\definecolor{c6-title-bkg}{HTML}{d3d5ed}
\definecolor{c6-title-text}{HTML}{220b1b}
\definecolor{c6-item-bkg}{HTML}{f5d8f1}
\definecolor{c6-item-text}{HTML}{220b1b}

\definecolor{c7-title-bkg}{HTML}{dad1ed}
\definecolor{c7-title-text}{HTML}{1c0b14}
\definecolor{c7-item-bkg}{HTML}{d58cb5}
\definecolor{c7-item-text}{HTML}{1c0b14}

\definecolor{c8-title-bkg}{HTML}{ded1ec}
\definecolor{c8-title-text}{HTML}{633273}
\definecolor{c8-item-bkg}{HTML}{D8BFD8}
\definecolor{c8-item-text}{HTML}{7c5997}

\definecolor{c9-title-bkg}{HTML}{e5d1eb}
\definecolor{c9-title-text}{HTML}{6c2f6b}
\definecolor{c9-item-bkg}{HTML}{f0e0f6}
\definecolor{c9-item-text}{HTML}{885591}

\definecolor{c10-title-bkg}{HTML}{ebd1e7}
\definecolor{c10-title-text}{HTML}{722e5f}
\definecolor{c10-item-bkg}{HTML}{f5e2f3}
\definecolor{c10-item-text}{HTML}{915487}

\definecolor{avg-title-bkg}{HTML}{f3f3f3}
\definecolor{avg-title-text}{HTML}{000000}
\definecolor{avg-item-bkg}{HTML}{f3f3f3}
\definecolor{avg-item-text}{HTML}{000000}

\newcommand{\addpadding}{%
  \rule{0pt}{\dimexpr\normalbaselineskip-0.5pt\relax}%
}

\newcommand{\ct}[2][c0]{\addpadding{\cellcolor{#1-item-bkg}\textcolor{#1-title-text}{#2}}}

\ExplSyntaxOn

\cs_generate_variant:Nn \coffin_typeset:Nnnnn {Nffff}

\NewDocumentCommand\rotbox{ O{l,H} D<>{0pt,0pt} m m}{
    \hcoffin_set:Nn \l_tmpa_coffin {#4}
    \coffin_rotate:Nn \l_tmpa_coffin {#3}
    \coffin_typeset:Nffff \l_tmpa_coffin 
        {\clist_item:nn{#1}{1}}
        {\clist_item:nn{#1}{2}}
        {\clist_item:nn{#2}{1}}
        {\clist_item:nn{#2}{2}}
}
\ExplSyntaxOff

\newlength{\ccustomlen}
\newcommand{\ccustom}[3][c0]{%
    \cellcolor{#1-item-bkg}{%
        \rotbox[l,t]{90}{%
            \parbox[t]{\ccustomlen}{%
                \ifthenelse{\isempty{#3}}{%
                    \mbox{%
                        \kindatiny\textcolor{#1-title-text}{#2}%
                    }%
                }{%
                    \kindatiny\textcolor{#1-title-text}{#2} \\%
                    \tiny{\textcolor{#1-item-text}{\it #3}}%
                }%
            }%
        }%
    }%
}

\newcommand{\cb}[3][c0]{%
    \setlength{\ccustomlen}{1.5cm}%
    \ccustom[#1]{#2}{#3}%
}

\usepackage{xspace}

\makeatother
\title{The Open Polymers 2026 (\op{}) Dataset and Evaluations }
\author[1, *, \dagger]{Daniel S. Levine}
\author[2, *]{Nicholas Liesen}
\author[3]{Lauren Chua}
\author[2]{James Diffenderfer}
\author[2]{Helgi I. Ingolfsson}
\author[2]{Matthew P. Kroonblawd}
\author[3]{Nitesh Kumar}
\author[2]{Amitesh Maiti}
\author[2]{Supun S. Mohottalalage}
\author[1]{Muhammed Shuaibi}
\author[2]{Brian Van Essen}
\author[1]{Brandon M. Wood}
\author[1]{C. Lawrence Zitnick}
\author[3, \dagger]{Samuel M. Blau}
\author[2, \dagger]{Evan R. Antoniuk}

\affiliation[1]{FAIR at Meta}
\affiliation[2]{Lawrence Livermore National Laboratory}
\affiliation[3]{Lawrence Berkeley National Laboratory}
\contribution[*]{Co-first Author}
\contribution[\dagger]{Co-corresponding Author}

\abstract{Polymers constitute the molecular foundation of living organisms and their synthetic counterparts drive transformative advances across medicine, consumer products, and energy technologies. While machine learning (ML) models have been trained on millions of quantum chemical atomistic simulations for materials and/or small molecular structures to enable efficient, accurate, and transferable predictions of chemical properties, polymers have largely not been included in prior datasets due to the computational expense of high quality electronic structure calculations on representative polymeric structures. We address this shortcoming with the creation of the Open Polymers 2026 (\op{}) dataset, which contains more than 6.35 million density functional theory (DFT) calculations on up to 360 atom clusters derived from polymeric systems, comprising over 1.2 billion total atoms. \op{} captures the chemical diversity that makes polymers intrinsically tunable and versatile materials, encompassing variations in monomer composition, degree of polymerization, chain architectures, and solvation environments. \op{} is directly compatible with the Open Molecules 2025 dataset, and we show that augmenting ML model training with \op{} improves model performance for polymer prediction tasks, and in particular for reactive configurations. We also publicly release the \op{} dataset to help further the development of ML models for polymers, and more broadly, strive towards universal atomistic ML models.}

\metadata[Dataset Updated with \op{}]{\url{https://huggingface.co/facebook/OMol25}}
\metadata[Code]{\url{https://github.com/facebookresearch/fairchem}}
\correspondence{D.S.L. (\email{levineds@meta.com}), S.M.B. (\email{smblau@lbl.gov}), E.R.A. (\email{antoniuk1@llnl.gov})}

\let\oldaddcontentsline\addcontentsline
\renewcommand{\addcontentsline}[3]{}

\mciteErrorOnUnknownfalse

\begin{document}
\maketitle

\section{Introduction}
Polymers are ubiquitous in modern life, employed in diverse areas from common consumer packaging to cutting-edge technologies, such as additive manufacturing and lithium-ion batteries. Simultaneously, the stability and tunability that makes polymers so useful has also placed them at the forefront of societal issues including the accumulation of plastic waste and the prevalence of microplastics \cite{jambeck_plastic_2015,li_potential_2023}. As a result, there is an urgent need to innovate in polymer design and synthesis, such that next-generation polymeric materials can be rapidly developed to meet these emerging societal and technological challenges.

Polymers are composed of long chains of repeating chemical units (monomers) whose architecture and composition can be systematically varied, thereby providing an additional dimension of tuning material properties. However, understanding structure-property relationships in polymers is inherently complex since it is dependent on capturing both atomic-scale interactions such as covalent bonds, hydrogen bonding, and dispersion, as well as interactions between collections of polymer chains, such as chain entanglements \cite{song_high-performance_2020,xie_hydrogen_2021,chen_covalent_2024}. Despite these challenges, computational modeling has become an essential tool for understanding polymer chemistry, providing a detailed atomistic description of polymer structure and dynamics \cite{gartner_modeling_2019}. Predictive insight into polymer degradation processes could unlock new materials properties and recycling schemes, which are needed to address the ecological impact of polymer materials.

Polymer simulations leverage different computational methods depending on the length-scale and time-scale of interest. Density functional theory (DFT) calculations can provide a high-accuracy description of the electronic structure of monomers/oligomers of O(100) atoms \cite{aldeghi_graph_2022,melenkevitz_density_1991}. While DFT is employed to study the reactivity and optoelectronic properties of small oligomers, it is too computationally burdensome to simulate larger systems or polymer chain dynamics. Molecular dynamics (MD) simulations with classical force fields (FFs) have long been used to simulate the dynamics of polymers with typical length-scales up to 100~\AA{} and time-scales up to 100~ns \cite{gartner_modeling_2019,gooneie_review_2017,bishop_molecular_1979}. However, there are few options for classical FFs that offer broad coverage over common polymer functional groups. Rather, a number of hand-tuned classical FFs parameterized on a single polymer family and proprietary models have been used for decades to perform MD simulations on polymer systems, including COMPASS and PCFF \cite{sun_ab_1994,sun_compass_1998}. In practice, this gap in FF availability often leads to the adoption of simple classical FFs inherited from small-molecule organic condensed phase FFs like OPLS and Dreiding, but this demands further tuning and careful model validation \cite{mayo_dreiding_1990,jorgensen_development_1996,kaminski_evaluation_2001}. Further, these classical FFs cannot accurately describe reactivity and exhibit limited transferability to chemical environments and conditions that they were not explicitly trained on \cite{gartner_modeling_2019}.

Machine Learning Interatomic Potentials (MLIPs) have recently emerged as transformative tools for atomistic simulations that can fill in this distinct gap in the need for generalizable, reactive FFs for condensed-phase polymer simulations  \cite{wood2025uma,rhodes_orb-v3_2025,batatia_mace_2023}. When trained on large datasets of DFT calculations, MLIPs can provide near-DFT-level accuracy energy and force predictions, while being several orders of magnitude faster \cite{chiang_mlip_2025}. MLIPs thus have the potential to enable large-scale polymer molecular dynamics simulations over long time scales while faithfully describing atomistic interactions and reactivity. Recently, large-scale pre-trained MLIPs have been developed that train on DFT calculations covering multiple material types, including molecules, metal-organic frameworks, and crystalline materials \cite{wood2025uma}. While these MLIPs seek to be chemically universal, they do not include polymeric simulations in their training data, primarily due to the lack of suitable open-source polymer DFT datasets. Nevertheless, ensuring that MLIPs accurately generalize to polymers is essential for building a true atomistic foundation model \cite{yuan2025foundationmodelsatomisticsimulation}.

The lack of polymer DFT datasets for MLIP training can be partially attributed to the large computational expense required to sufficiently sample the vast chemical and structural space of polymer chemistries. While the recently released Open Molecules 2025 (OMol25) dataset \cite{levine2025openmolecules2025omol25} did include some larger polypeptide structures, in addition to small molecules, electrolytes, and metal complexes, it did not sample polymer structures more broadly. Meanwhile, classical polymer force field models are traditionally parameterized on a single polymer composition or narrow polymer family, such as polycarbonates \cite{sun_ab_1994}. Similarly, specialized polymer MLIP models have been trained on O(1000) polymer structures originating from \textit{ab initio} MD \cite{matsumura_generator_2025}. The recently published SimPoly effort reported a dataset for MLIP training consisting of DFT calculations on 680,000 polymer configurations \cite{simm_simpoly_2025}. These configurations were sampled from 130 unique polymer compositions and focus on stable homopolymers. While SimPoly was a major advance over previous work, it highlights the need for datasets encompassing more diverse polymer chemistries. 

\begin{figure}[h]
    \centering
    \includegraphics[width=1.00\linewidth]{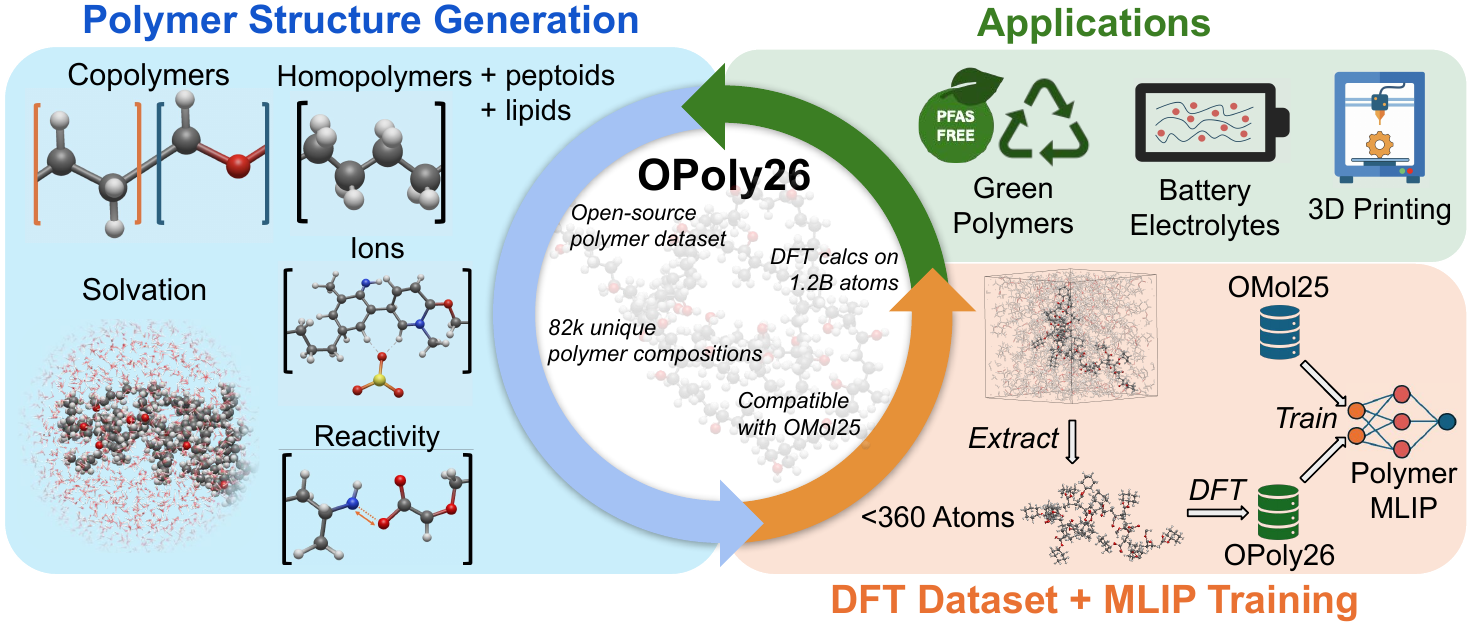}
    \caption{Overview of \op{}. We generate diverse polymeric structures that cover a broad spectrum of polymer chemistries including homopolymers and highly diverse copolymers, reactive configurations, and interactions between polymers and solvent molecules and ionic species. We then extract >6M substructures with <360 atoms, capping dangling bonds with hydrogens as necessary, and simulate each with DFT. We train MLIP models on both the \op{} and OMol25 datasets, which employed identical DFT settings, to ensure model generalization across molecular chemistry domains, unlocking improved simulation capabilities in a range of high-impact applications.}
    \label{fig:omers_overview}
\end{figure}

Here, we present the Open Polymers 2026 (\op{}) Dataset---a large-scale, openly available resource designed to enable the development of MLIPs for polymeric systems (Figure \ref{fig:omers_overview}). Diverse polymeric structures are obtained by simulating polymer structure dynamics, including molecular dynamics (MD) simulations of 94k unique amorphous polymer simulation cells for a cumulative total of over 239,000~ns. We extract substructures from these larger polymeric systems, capping dangling bonds with hydrogens as necessary, and form the total MLIP dataset from over 6.35 million DFT single-point calculations including 1.2B total atoms. \op{} was specifically designed to capture the vast diversity of chemical environments found in polymer materials, spanning various linear chain architectures, molecular weight distributions, solvation environments, condensed-phase reactions, and polymer-ion interactions. \op{} also used the identical DFT settings as OMol25, making it trivial to combine the datasets for MLIP training. We demonstrate that an MLIP trained on a combination of \op{} and OMol25 is more accurate for polymer systems than an MLIP trained on OMol25 alone without regressing on the performance for other molecular chemistry domains. \op{} in particular dramatically improves model performance for condensed phase reactive configurations. To encourage the open development of improved MLIP models, we provide all of the \op{} data with a CC-BY-4.0 license.

\section{Open Polymers 2026 Dataset}
\begin{figure}[h]
    \centering
    \includegraphics[width=1.00\linewidth]{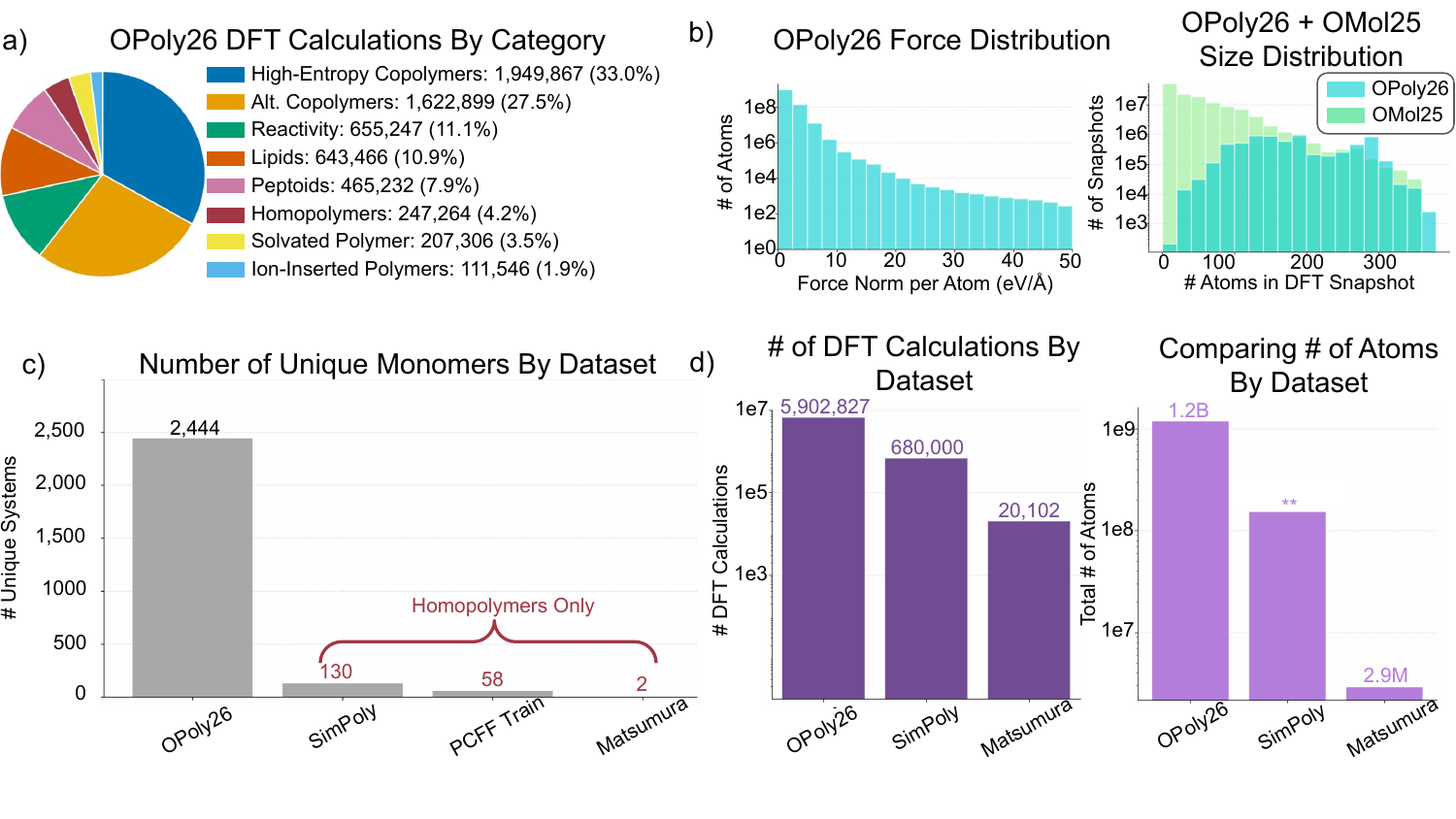}
    \caption{Summary of \op{} training dataset statistics. a) Distribution of \op{} DFT calculations by polymer categories (train split only). b) Distribution of force norms per atom across all \op{} DFT calculations (left) and distribution of the number of atoms in each \op{} DFT calculation in comparison to OMol25 (right). c) Comparison of the number of unique monomers across polymer datasets including SimPoly \cite{simm_simpoly_2025}, the original PCFF train dataset \cite{sun_ab_1994}, and the recent work of Matsumura et al. \cite{matsumura_generator_2025}. d) Number of atoms contained in polymer DFT datasets meant for MLIP training including the \op{} training split, the entire SimPoly dataset \cite{simm_simpoly_2025}, and the work of Matsumura et al. \cite{matsumura_generator_2025}. **As SimPoly is not yet publicly available, we estimate the total number of atoms by assuming the highest possible number of atoms in each DFT calculation.}
    \label{fig:omers_summary}
\end{figure}

In order to sample a broad range of polymer structures in \op{}, we include diverse monomeric repeat units, backbone structures, and supramolecular arrangements. In Section~\ref{ssec:poly_compositions}, we describe the various sources of monomers which span the field of polymer science, including optical polymers, fluorinated polymers, polymer electrolytes, and more. Those monomers are assembled into a range of polymer architectures (Section \ref{ssec:poly_architectures}), which were simulated with multiple techniques to generate diverse 3D structures for each polymer system (Section \ref{ssec:simulating_dynamics}-\ref{ssec:reactivity}). Finally, we extract a total of 6.35 million unique substructures from these larger polymer systems and perform high-throughput DFT calculations to generate the \op{} MLIP dataset (Figure \ref{fig:omers_summary}).

Altogether, the \op{} dataset is sourced from 2,444 unique monomers, with more than 1.2 billion total atoms across all the DFT calculations. The MD trajectories alone consist of 43,176 simulations cells of $\sim300$ atoms and 51,260 simulation cells of $\sim5000$ atoms, simulated over a cumulative 239,000~ns. These polymer systems include various linear chain architectures, solvation environments, chain lengths, and comonomer ratios, thereby capturing the structural complexities and diversity of atomistic interactions in polymers more comprehensively than prior homopolymer-only datasets. The \op{} dataset contains more than 6.35 million DFT calculations and required $\sim1.2$B CPU core-hours.

\subsection{Polymer Compositions} \label{ssec:poly_compositions}
Polymers are an extremely versatile class of materials that can include a diverse range of chemistries in their repeat units. In the following subsections, we outline the categories of polymer compositions in \op{} and how they were sampled. Our detailed procedure for obtaining all polymer compositions is available in Appendix \ref{sec:app:polymer_compositions}.

\subsubsection{Traditional Polymers}\label{ssec:trad_polymers} 
For a representative set of commonly used synthetic polymers, we leveraged the collection of homopolymers from the RadonPy benchmark set \cite{hayashi_radonpy_2022}. Notably, these polymers have the largest number of experimental measurements in the PolyInfo database \cite{otsuka_polyinfo_2011}, indicating that they are widely used and relevant materials. We further augmented this set with hand-picked polymer compositions from textbooks to broaden the distribution of represented polymers \cite{bicerano_prediction_2002}. The final dataset contains a diverse range of polymer backbones including polystyrenes, polyacrylates, and polyurethanes. A detailed breakdown is provided in Appendix \ref{secsec:app:trad_polymers}.

\subsubsection{Fluoropolymers}\label{ssec:fluoro_polymers}
Fluoropolymers are a critical class of polymers that fall within the broader class of materials known as polyfluoroalkyl substances (PFAS). PFAS materials have recently gained significant attention due to their widespread usage in consumer products, coupled with their extreme persistence in the human body and environment as well as significant human health risks \cite{lohmann_PFAS_2020,ebiomedicine_forever_2023}. Due to the lack of open datasets of experimentally synthesized fluoropolymers, we extract fluoropolymer compositions from the Open Macromolecular Genome (OMG) of ML-generated polymer systems that are compatible with known polymerization reactions \cite{kim_OMG_2023}. To ensure a diverse range of fluoropolymer chemistries are represented, we evenly sample fluorinated compositions from the 17 polymerization reactions described in OMG.

\subsubsection{Optical Polymers}\label{ssec:opt_polymers}
Polymers with $\pi$-conjugated backbones possess unique optical and electronic properties \cite{rasmussen_conjugated_2023}. As a result, conjugated polymers are important to a wide range of innovative optical technologies including photocatalysts, organic solar cells, photovoltaics, and organic light-emitting diodes \cite{bai_accelerated_2019,zhang_thiadiazole-based_2021,xu_new_2021}. We sample optical polymer systems from the Polymer Chemprop dataset \cite{aldeghi_graph_2022}, which spans the space of 892 monomers originally defined by Bai et al. \cite{bai_accelerated_2019}. 

\subsubsection{Polymer Electrolytes}\label{ssec:electro_polymers}
Solid-phase polymer electrolytes are being widely explored as replacements to currently used liquid electrolytes in lithium-ion batteries due to their improved safety (i.e. lower volatility and flammability) coupled with their ease of thin-film processing \cite{song_reflection_2023,xie_accelerating_2022,lei2023self,yang2023novo}. In \op{}, we include 300 diverse polymer compositions (without ions) from the PolyGen dataset of amorphous polymer electrolytes \cite{xie_accelerating_2022,lei2023self,yang2023novo}. To ensure that we appropriately capture the interactions between polymer chains and ionic species, we also create ion-inserted polymer systems, described in Appendix \ref{secsec:app:ion_inserted_polymers}

\subsubsection{Peptoids}
Peptoids are synthetic biopolymers that consist of N-substituted glycines. Importantly, this chemical structure of peptoids allows for a wide array of side-chain chemistries to be incorporated into the peptoid structure, thereby allowing for the systematic design and tailoring of material properties. Our catalog of peptoid monomers is comprised of experimentally synthesized, charge-neutral peptoid monomers from the Peptoid Data Bank \cite{eastwood_guidelines_2023}, as well as hand-collected peptoid monomers that are of particular interest for peptoid-based EUV lithography photoresists \cite{acsmacrolett.5c00320}.

\subsubsection{Lipids}
Lipids, although not polymeric in nature, are included in \op{} because lipid simulations suffer from many of the same challenges as polymer simulations. Lipids are a larger class of hydrophobic/amphipathic biomolecules that are insoluble in water. Many lipids spontaneously form supramolecular assemblies, such as bilayers, in aqueous environments, and the material properties of these superstructures are defined by the specific lipid composition and detailed environmental conditions \cite{vattulainen_lipid_2011}. Capturing lipid bilayer properties in simulations can be challenging and often requires integrating over several length- and time-scales \cite{vattulainen_2019,sansom_2019}. To span a diverse set of common lipids in different bilayer environments, we sample lipid simulation snapshots from the NMRlipid database \cite{samuli_2024}. 

\subsection{Polymer Architectures} \label{ssec:poly_architectures}
We strive to maximize structural and interaction diversity spanned by \op{} to enhance the generalizability of MLIPs trained on the dataset while ensuring that polymer structures are computationally feasible to simulate. As outlined in Table \ref{tab:dataset}, we generate bulk, amorphous cells of the polymeric structures with the RadonPy package \cite{hayashi_radonpy_2022}, where chain architectures are limited to linear homopolymers and copolymers (including alternating and random). Additionally, we perform MD simulations of alternating copolymers in the presence of 17 different explicit solvent environments (Appendix \ref{sssec:app:solvated_cells}). To increase structural and interaction diversity, we also generate high-entropy copolymers, which are random copolymers consisting of between 4-10 distinct monomers, where the monomers are sampled from all polymer compositions in Sections \ref{ssec:trad_polymers}-\ref{ssec:electro_polymers}. Peptoids are sampled in a similar manner but with a small amount of water also present. 

\subsection{Simulating Polymer Dynamics} \label{ssec:simulating_dynamics}
\begin{figure}[H]
    \centering
    \includegraphics[width=1.00\linewidth]{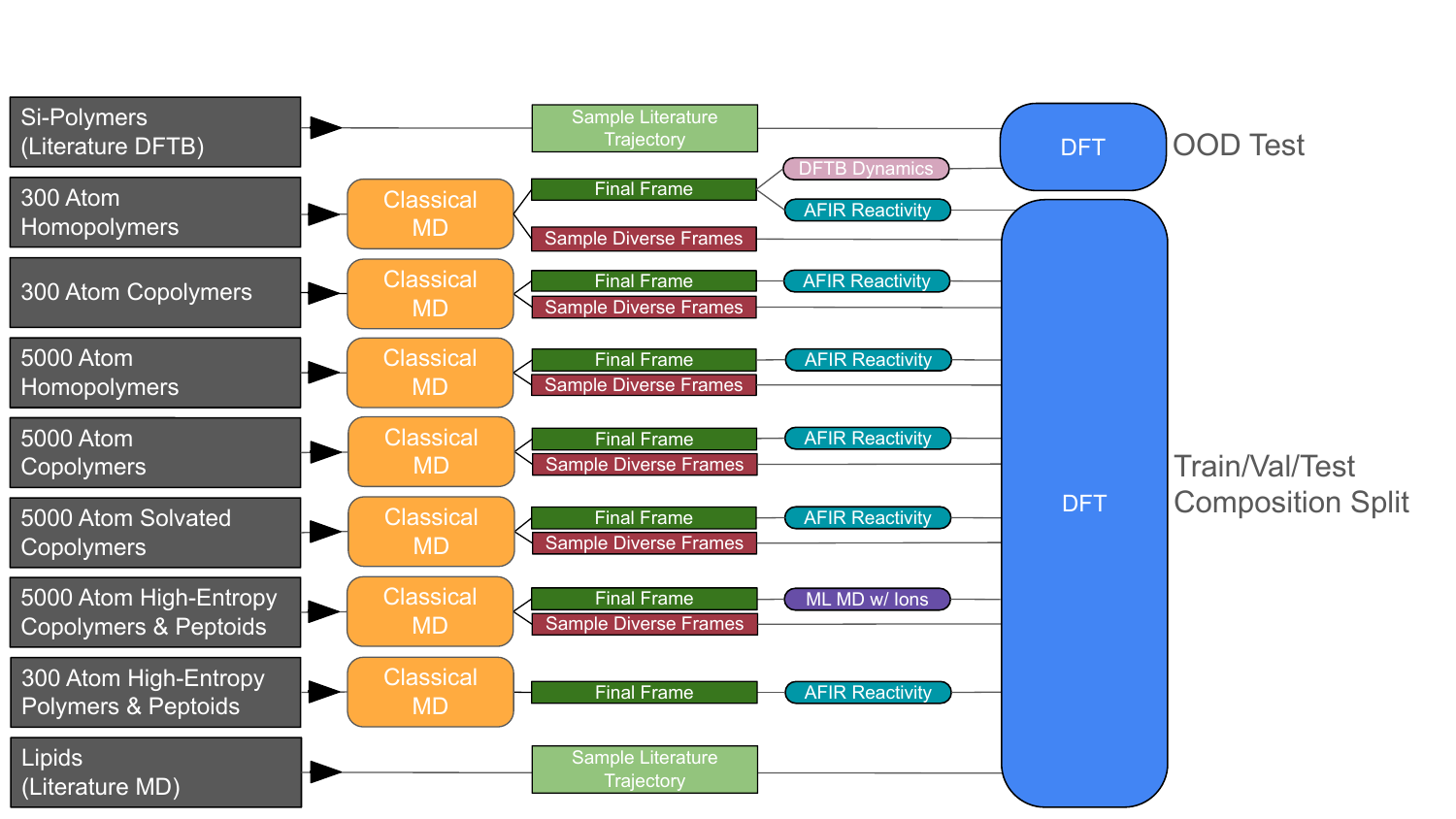}
    \caption{Overview of the \op{} data generation process. Starting from a diverse collection of polymeric compositions (Section \ref{ssec:poly_compositions}), we apply a sequence of computational steps tailored to each class of materials. In general, we first use classical MD simulations to generate a broad distribution of polymer configurations (Section \ref{ssec:simulating_dynamics}). From these trajectories, we sample both the relaxed final frame (for use in further calculations) as well as diverse frames from across the MD trajectory, as described in Section \ref{sec:extracting_substructures}. All resulting structures are subsequently passed into a final DFT calculation, yielding the MLIP training, validation, and test sets split by composition. Two additional out-of-distribution test sets are generated from DFTB dynamics and silicone polymer structures.}
    \label{fig:polymer-structures}
\end{figure}

The combination of the polymer compositions (Section \ref{ssec:poly_compositions}) with the polymer architectures (Section \ref{ssec:poly_architectures}) results in the creation of 94k unique amorphous polymer simulation cells (see Table \ref{tab:dataset}). Next, we employ a variety of simulation strategies, including MD, MLIP-MD, artificial force induced reactivity (AFIR), and density functional tight binding (DFTB), to yield a rich dataset of polymer conformations for MLIP training and testing (Figure \ref{fig:polymer-structures}).

\begin{table}[H]
\caption{Summary of input structures to the \op{} MD-based data generation. }
\centering
\resizebox{1.0\textwidth}{!}{%
\begin{tabular}{cccl}
\toprule
\multicolumn{1}{l}{} & & \# MD & \\ 
 Category Name & Compositions & Trajectories & Cell Description \\ \midrule 
300 Atom Homopolymers & Fluoro, Trad., Optical, Electrolytes     & 2,155 & 3 chains of 100 atom homopolymers \\
5000 Atom Homopolymers & Fluoro, Trad., Optical, Electrolytes      & 2,046   & 10 chains of 500 atom homopolymers \\ \midrule
300 Atom Copolymers  & Fluoro, Trad., Optical, Electrolytes    & 17,017   & 2 chains of 150 atom alternating copolymers \\
5000 Atom Copolymers & Fluoro, Trad., Optical, Electrolytes   & 16,247     & 10 chains of 500 atom alternating copolymers    \\
300 Atom High-Entropy Copolymers & High-Entropy Copolymers  & 17,434    & 3 chains of 100 atom random polymers with between 3-6 unique comonomers              \\
5000 Atom High-Entropy Copolymers& High-Entropy Copolymers  & 19,622    & 10 chains of 500 atom random polymers with between 5-10 unique comonomers               \\ \midrule
5000 Atom Solvated Copolymers             & Fluoro, Trad., Optical, Electrolytes     & 8,383   & 1 chain of 500 atom random copolymer with 4500 solvent atoms \\ \midrule
300 Atom Peptoids & Peptoids   & 6,570    &   3 chains of 100 atom high-entropy copolymers with 10 water molecules      \\
3000 Atom Peptoids & Peptoids   & 4,962   &   10 chains of 300 atom high-entropy copolymers with 100 water molecules \\
\bottomrule
\end{tabular}%
}
\label{tab:dataset}
\end{table}

To generate large quantities of polymer structures, we carried out a series of classical all-atom MD simulations using the LAMMPS MD suite \cite{thompson2022lammps} and the RadonPy tools made available by Hayashi et al \cite{hayashi_radonpy_2022}. Additional details of our MD procedure are provided in Appendix \ref{sec:app:md_details}. The final frames of high-entropy copolymers and peptoids further had ions inserted into voids and were then subjected to MLIP-MD simulations. Additional details are provided in Appendix \ref{sec:app:ion_inserted_mlip_md}. Finally, homopolymer final frames were subjected to density functional tight binding (DFTB) simulations to generate a set of distinct configurations for an out-of-distribution test set. Additional details are provided in Appendix \ref{sec:app:dftb}.

\subsection{Reactivity} \label{ssec:reactivity}
To probe polymer-specific reactivity in realistic, complex condensed-phase environments, we generate reactive configurations through single-ended artificial force-induced reaction (AFIR) \cite{maeda2016afir, levine_large_2023} searches that target bond dissociation events within polymer chains. AFIR trajectories, which serve to approximate the minimum energy pathway between reactants and their products, are initialized from a converged MD trajectory frame resulting from Section~\ref{ssec:simulating_dynamics}, but additionally with a subset of structures subjected to protonation or deprotonation at an allowable functional group near the bond being stretched. Bond dissociations in polymer systems are important for the study of thermal, photochemical, and mechanical degradation as well as reversible and living polymerizations, but also serve as a facile way to explore polymer reactivity across various monomer and co-monomer compositions. We further note that the bond is intentionally stretched far beyond its breakage distance, often causing subsequent reactive events between dangling atoms and other polymer and/or solvent atoms. Additional details are provided in Appendix \ref{sec:app:react_details}.

\subsection{Extracting Polymer Substructures from Trajectories}
\label{sec:extracting_substructures}
Due to the large computational expense of DFT calculations, it is infeasible to perform single point DFT calculations on entire 5000 atom polymer simulation cells. Instead, we extract smaller substructures (<360 atoms) from the 5000 atom polymer simulation cells to capture a diversity of local chemical environments. From each MD trajectory, we intentionally sample a high proportion of frames from the simulated annealing portion, resulting in preferential sampling of non-equilibrium polymer structures (as seen in Figure \ref{fig:omers_summary} force distribution). To further aid in sampling diverse configurations within each trajectory, we sample frames of the trajectory with the largest dissimilarity to all other structures in the trajectory. Finally, structures suitable for molecular DFT are created by extracting hydrogen-capped shells of atoms, terminating between polymer repeat units. Our detailed procedure for extracting lipid and polymeric substructures is detailed in Appendix \ref{sec:app:extract}.

\subsection{DFT Calculation Details}
All DFT calculations are performed at the level of theory described in full detail in OMol25 \cite{levine2025openmolecules2025omol25}. Specifically, we employ the triple-zeta def2-TZVPD basis set \cite{weigend_balanced_2005} and the range-separated hybrid meta-GGA DFT functional $\omega$B97M-V \cite{mardirossian_b97m-v_2016} due to its performance as one of the most accurate functionals across various DFT functional benchmarks \cite{mardirossian_thirty_2017}. Complete details are provided in Appendix \ref{sec:app:dft_calcs}.

\subsection{\op{} Training, Validation, and Test Splits}
\op{} is divided into consistent training, validation, and test splits consisting of 5,902,827, 201,865, and 248,391 polymer substructures, respectively, based on the atomic composition of each datapoint. In other words, a composition which appears in one split does not appear in either of the other two. We further construct two additional out-of-distribution (OOD) test sets. One uses DFTB generated structures as detailed in Appendix \ref{sec:app:dftb}. Notably, this DFTB test set is OOD with respect to train/validation/test splits in terms of both the simulation method used and the temperature of the simulations (600~K). Another OOD test set is derived from a dataset of silicone polymers undergoing radiation-induced degradation processes \cite{kroonblawd_polymer_2022}. While this Si-polymers test set cannot be used with MLIP models trained on \op{} alone, as there is no silicon in the training data, it can be used with models trained on OMol25 or on both \op{} and OMol25. 

\subsection{Dataset Limitations}
With more than 6 million high-accuracy DFT calculations, \op{} is the largest and most comprehensive dataset for training MLIP models on non-biological polymers to date. Nevertheless, given the broad usage of the term ``polymer'', it is important to specifically outline classes of polymeric materials that are not explicitly covered by \op{}. First, \op{} does not contain gradient copolymers, block copolymers, branched polymers, crosslinked polymers or graft polymers, as these polymer chain architectures cannot be appropriately represented in our <360 atom DFT simulation cells. Nevertheless, given the common local chemical environments between these unseen polymer architectures and those present in \op{}, future work will explore how MLIPs trained on \op{} can generalize to these out-of-distribution polymer architectures. We also note that \op{} does not include any silicon or other heavy p-block element-containing polymers. 
%Although several force fields contain atom types relevant specifically for polysiloxane structures,\cite{sun_ab_1994,sun_compass_1998} we avoided their usage as we expect that these force fields will not transfer well to the broad polymer compositions in \op{}. However, the release of \op{} unlocks a new possibility to simulate silicon-containing polymers by jointly training on \op{} and large small-molecule datasets of Si-containing molecules, such as OMol25.\cite{levine2025openmolecules2025omol25} 
%Finally, to the extent that proteins can be considered biological polymers, we do not explicitly include protein structures in \op{}, although they are thoroughly sampled in OMol25. 
Finally, although all monomer compositions have been verified for chemical validity, not all monomer compositions nor polymer architectures present in \op{} are guaranteed to be synthesizable (see Appendix \ref{sec:app:polymer_compositions}).      

\section{Evaluations}

\begin{figure}[h]
    \centering
    \includegraphics[width=0.80\linewidth]{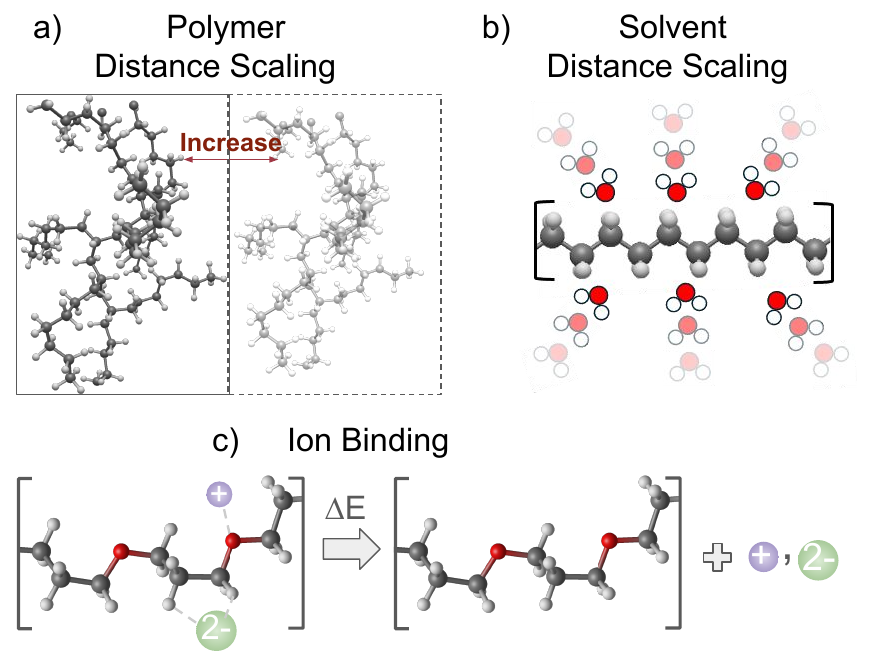}
    \caption{\op{} evaluation tasks. a) Interaction energy between two identical copies of a polymer chain is calculated as a function of the interchain separation distance. b) Solvent interaction energy is calculated by systematically increasing the distance between the chain and a shell of solvent molecules. c) Ion binding energy is calculated as the energy difference between the polymer-ion complex and the isolated ions and polymer.}
    \label{fig:evals}
\end{figure}

%We are in the process of finalizing a set of evaluation tasks that are designed to assess the performance of models trained on \op{} to accurately capture local polymer structures and their interactions with nearby chemical moieties. This manuscript will be updated with additional details and results on these evaluations in the near future.
Here, we define a set of polymer-specific evaluation tasks that are designed to assess the performance of models trained on \op{} to accurately capture interactions with nearby chemical moieties (see Figure \ref{fig:evals}). Future work will explore how well models trained on \op{} can capture local polymer structure and experimentally measured properties of higher molecular weight polymeric systems.

%\subsection{Conformers}
%The overall conformation of a polymeric material depends on many complex factors including the steric bulk of the monomers, stiffness of the backbone, torsional angle between monomer units, and interchain hydrogen bonding or $\pi$-interactions.\textcolor{red}{CITE} Taken together, all of these factors influence how tightly polymer chains can pack, which in turn influences nearly all of the downstream properties of the polymer. Indeed, the importance of accurately capturing polymer conformations is reflected in the fact that polymer-specific FF models are often benchmarked against how well they predict torsional potential energy profiles or conformational distributions. \textcolor{red}{CITE} Here, we perform polymer conformation evaluations analogous to those performed in OMol25. \cite{levine2025openmolecules2025omol25} Specifically, we compare the polymer conformation obtained when relaxing the structures with either DFT or an MLIP. \textcolor{red}{update with new method}
\subsection{Polymer Distance Scaling}
Accurately capturing interchain interactions in polymers is crucial, as even small changes to the polymer repeat unit has been shown to significantly impact their mechanical properties, phase transitions, and optical properties \cite{ishizaka_minor_2024,beljonne_interchain_2000,gonzalez_cortes_effect_2023}. It is particularly important to assess interchain interactions because they arise between long, entangled macromolecules and are therefore not represented in prior small-molecule MLIP training datasets. To this end, we perform an additional 8,230 MD simulations of random copolymers (10 chains of approximately 500 atoms) of unseen traditional polymer compositions, following the methodology in Appendix \ref{sec:app:md_details}. Isolated polymer chains are then extracted from the final frame of these MD trajectories. For each isolated polymer chain, we computed a periodic box which maximized contacts with its image, and then incrementally increased the cell size, thereby decreasing the interaction energy between the two chains. Up to 250 atoms were permitted and the chains were separated in increments of 0.25*the interchain centroid distance until the two chains were 5\AA{} apart. The difference in energy/force from the unseparated systems to each separation distance is compared to the equivalent value calculated with DFT, yielding an interaction energy/force MAE evaluation.

\subsection{Solvent Distance Scaling}
Understanding the solubility of a polymer in solution has a rich history of being studied through theoretical approaches such as Flory-Huggins theory, thermodynamic parameters such as $\chi$ and Hansen solubility parameters, and simulation approaches based on MD or coarse-grained simulations \cite{flory_thermodynamics_1941,huggins_theory_1942,hansen_universality_1969,ethier_predicting_2024,kozuch_predicting_2016}. Much of the motivation for seeking to understand polymer solubility stems from its importance in polymer processing, as polymers must often be dissolved in a suitable solvent before they can be processed \cite{ethier_predicting_2024}. To probe the ability of models trained on \op{} to accurately describe polymer-solvent interactions, we perform additional classical MD simulations on solvated traditional homopolymer compositions, as described in Appendix \ref{sssec:app:solvated_cells}. These simulations were performed with five out-of-distribution solvents not found in the train set of either \op{} or OMol25 (\ref{sec:app:solvents}). Following MD equilibration, we define a subsystem comprising the polymer chain and at least four solvent molecules nearest to the polymer chain (up to a maximum of 300 atoms) that lay within 3.5\AA{} of the polymer chain (this was reduced to 2.5\AA{} if too many solvent molecules were included at 3.5\AA{}). We perform DFT calculations on this subsystem, systematically increasing the pairwise distances between the solvent molecules and their separation from the polymer chain. Distances between the center of mass of the polymer chain and of each solvent molecule were dilated by a factor of 1.1, 1.3, 1.5, and 1.8. We then compare the total energy/forces of these systems calculated with DFT and the MLIP vs. the total energy/forces of the undilated system to yield an interaction energy/force MAE evaluation metric.

\subsection{Ion Binding}
The interaction between ions and polymers underpin many critical technologies. For instance, lithium-ion coordination in solid state polymer electrolytes impacts the ionic conductivity of lithium-ion batteries, and the CO$_2$ sorption performance of ionic polymer carbon capture materials can be modulated through the choice of ionic functional groups \cite{allam_molecular_2025,da_luz_polyionic_2022}. We construct a polymer–ion binding evaluation task by inserting up to five randomly selected polyatomic ions not included in the \op{} training set (see Table \ref{tab:ions}) into each of 1,102 3000-atom MD-equilibrated peptoid structures, which are then subjected to 2000 steps of MLIP-MD to allow the polymer backbone to reorient due to the presence of the ions(Appendix \ref{sec:app:ion_inserted_mlip_md}). The energy/force difference between the ion-polymer complex and the isolated ion and isolated polymer chain can then be calculated with DFT and the MLIP, yielding an interaction energy/force MAE.
\section{Baseline Models}
To establish a set of baseline results, we evaluate the performance of eSEN models \cite{fu_learning_2025} trained on OMol25, \op{}, and OMol25 + \op{}. We note that the primary focus of this work is not to comprehensively evaluate MLIP architectures or training procedures, but to explore how training on different combinations of small molecule and polymer data improves the downstream model performance in these different domains. All three eSEN models are trained for 12 epochs. As an additional point of comparison, we evaluate the performance of UMA-s-1p1 (no \op{} in training) and UMA-s-1p2 (includes \op{} in training) \cite{wood2025uma}. These two UMA models were not trained in a step-equivalent manner with each other or the eSEN models and therefore should not be interpreted as a direct data ablation experiment. There are also non-trivial architectural changes between UMA-s-1p1 and UMA-s-1p2 \cite{wood2025uma}. 
\section{Results}

\subsection{Energy and Force Test Sets}
%A core question that we wish to answer with \op{} is if existing state-of-the-art MLIP models that have been trained entirely on small-molecule data can generalize to larger polymeric structures, consisting of similar chemical motifs. This question has important consequences for developing universal MLIP models, as it would demonstrate if MLIP models can generalize to arbitrarily large chemical systems as long as similar local chemical environments have been seen in training. 
By comparing \op{} test set performance of models trained on only OMol25, only \op{}, and OMol25 + \op{}, we seek to probe the benefit of polymer-specific training data for polymer-specific MLIP prediction tasks --- and whether the addition of polymer data degrades performance in other molecular domains. 
%To explore this hypothesis, we evaluate our models' performance on predicting the energy and per-atom forces of structures from both \op{} and OMol25. Similarly, we train eSEN models on i) \op{} only, ii) OMol25 only, and iii) both \op{} and OMol25 train splits. 
%All models are evaluated with total energy and atomic force mean absolute error (MAE). 

\begin{table}[h!]
\caption{Summary of energy and force prediction performance on \op{} test sets for models trained on different training datasets. For each model, we report the mean absolute error in total energy (meV) and forces (meV/\AA). For the compositional test set, models trained on \op{} have substantially reduced energy error compared with a model trained only on OMol25. All models show a moderate improvement on the DFTB test set when trained on \op{} data. The Si-polymer test set has substantially larger energy errors, and while the addition of \op{} data does meaningfully reduce energy error, the reduction is not as substantial as for the test composition set. }
\begin{threeparttable}
\centering
\resizebox{\textwidth}{!}{%
\begin{tabular}{l SS SS SS}
 &
\multicolumn{2}{c}{\ct[c3]{\it Test Composition}} &
\multicolumn{2}{c}{\ct[c5]{\it DFTB}} &
\multicolumn{2}{c}{\ct[c6]{\it Si-Polymers}}
\\
%\cmidrule(l){3-10} 
 Model & \mcc{\ct[c3]{Energy MAE$\downarrow$}}   & \mcc{\ct[c3]{Force MAE$\downarrow$}}     
 & \mcc{\ct[c5]{Energy MAE$\downarrow$}}   & \mcc{\ct[c5]{Force MAE$\downarrow$}}  
 & \mcc{\ct[c6]{Energy MAE$\downarrow$}}   & \mcc{\ct[c6]{Force MAE$\downarrow$}}  
 \\ 
 \midrule[1pt]
OMol25 Only
& 70.5 & 6.2 & 37.8 & 4.4 & 190.7 & 6.1 \\
\midrule
UMA-s-1p1
& 73.7 & 6.9 & 47.3 & 5.2 & 234.4 & 8.4 \\
\midrule[2pt]

\op{} Only 
 & 17.7 & 3.7 &  26.9 & 3.7 & {{---}} & {{---}}\\
\midrule

\op{} + OMol25
 & 24.3 & 4.4 & 32.1 & 4.0 & 174.2 & 6.1\\
\midrule

UMA-s-1p2
& 19.0 & 3.8 & 27.0 & 3.6 & 181.1 & 6.5 \\

\bottomrule

\end{tabular}%
}
\begin{tablenotes}
  \item \kindatiny Energy MAE(meV), Force MAE(meV/\angs)
\end{tablenotes}
\end{threeparttable}
\label{tab:test_omers_ood_total}
\end{table}

We find that polymer-specific data is necessary to obtain sub-kcal/mol total energy accuracy (``chemical accuracy'' = 1 kcal/mol = 43 meV) on diverse polymer structures (Table \ref{tab:test_omers_ood_total}). Specifically, the \op{} and \op{} + OMol25 models show more than a  $\sim$66\% reduction in the total energy mean absolute error (MAE) on the compositional test split compared to a model trained only on OMol25. Similarly, atomic force MAE shows a significant improvement, although the force errors on the \op{} splits for all models are fairly small, on average. To go beyond broadly averaged insight, we further decompose the compositional test set errors into different chemical categories in the following section. \op{} data also marginally improves model performance on the DFTB test set, although the OMol25-only model errors were already fairly low, perhaps because the DFTB test set samples solely homopolymers. Finally, the Si-polymer test set is substantially more challenging, featuring high-energy systems far from equilibrium generated under simulated irradiation conditions of silicone polymers. Given that neither simulated irradiated systems nor silicone polymers are present in the \op{} training set or OMol25, it is unsurprising that energy errors on this test set are significantly higher across models compared with the previous two test sets (though still low enough to likely be practically useful for simulations). We note that the \op{} + OMol25 model has $\sim$9\% lower energy error compared with the OMol25-only model, demonstrating that organic polymer data may hold some value for silicon-based polymers. %Finally, all five models achieve low errors on the DFTB test set, perhaps because only the DFTB set samples solely homopolymers. 
%Across all test sets in Table \ref{tab:test_omers_ood_total}, we find significantly lower energy and force MAE values for the models that include \op{} in training, highlighting the importance of training on polymeric systems. 
%That both \op{}-trained models performed very similarly despite the fact that OMol25 contains so much more data is encouraging that the \op{} data are truly complementary to the existing data in OMol25; there is no friction from training similar data. All models perform similarly well on the DFTB and Si-polymers test sets.

\newcolumntype{Y}{>{\centering\arraybackslash}X}
\begin{table}[htbp]
\centering
\caption{Breakdown of energy prediction errors on the OPoly26 composition test split. For each model, we report the mean absolute error in total energy (meV) on various polymer types, as defined in Figure \ref{fig:omers_summary}. The number below each polymer type indicates the number of DFT snapshots corresponding to that polymer class in the test set. A breakdown of force errors is given in Table \ref{tab:test_force_errors}.}
\label{tab:test_energy_errors}
\scriptsize 
\setlength{\tabcolsep}{2pt} % Slightly tighter for very wide tables
\begin{tabularx}{\textwidth}{@{} l *{8}{Y} @{}}
\textbf{Model} 
& \ct[c3]{\makecell[b]{\textbf{High Entropy} \\ \textbf{Copolymers} \\ (N=115,866)}}
& \ct[c4]{\makecell[b]{\textbf{Alt.} \\ \textbf{Copolymers} \\ (N=36,592)}}
& \ct[c5]{\makecell[b]{\textbf{Lipids} \\ (N=8,676)}}
& \ct[c6]{\makecell[b]{\textbf{Peptoids} \\ (N=47,595)}}
& \ct[c3]{\makecell[b]{\textbf{Reactivity} \\ (N=18,462)}}
& \ct[c4]{\makecell[b]{\textbf{Homopolymers} \\ (N=5,702)}}
& \ct[c5]{\makecell[b]{\textbf{Solvated} \\ \textbf{Polymers} \\ (N=8,878)}}
& \ct[c6]{\makecell[b]{\textbf{Ion-Inserted} \\ \textbf{Polymers} \\ (N=6,620)}} \\
\midrule
OMol25 Only 
& 16.4 & 18.0 & 25.3 & 16.3 & 721.7 & 14.3 & 18.3 & 59.7 \\
\midrule
UMA-s-1p1 
& 16.9 & 19.3 & 30.2 & 18.2 & 753.6 & 16.1 & 18.0 & 54.9 \\
\midrule[2pt]
OPoly26 Only 
& 7.3 & 6.7 & 20.4 & 7.4 & 122.7 & 5.9 & 8.9 & 59.6 \\\midrule
OPoly26 + OMol25 
& 11.3 & 10.7 & 17.5 & 10.9 & 171.3 & 10.2 & 13.0 & 49.8 \\
\midrule
UMA-s-1p2 
& 9.1 & 8.8 & 11.1 & 8.8 & 134.9 & 8.2 & 9.7 & 29.3 \\
\bottomrule
\end{tabularx}
\begin{tablenotes}
  \item \kindatiny Energy MAE(meV)
\end{tablenotes}
\end{table}

To better understand how model performance improvement from \op{} data maps onto different polymer domains, in Table \ref{tab:test_energy_errors} we separate compositional test set energy prediction errors by chemical category. The equivalent breakdown of force errors is provided in Table \ref{tab:test_force_errors}. Although we observe a reduction in energy errors across all polymer types, the largest reduction in error by far is for configurations from the Reactivity portion of the dataset. Indeed, the 76\% improvement from 721.7 meV for the OMol25-only model to 171 meV for the \op{} + OMol25 model is the difference between being highly error-prone vs. being practical useful for simulations, providing immense scientific utility. In contrast, the OMol25-only model already achieves average energy prediction errors that fall within the threshold of chemical accuracy for non-reactive copolymers, lipids, homopolymers, solvated polymers, and peptoids, even without explicit training on polymeric structures. Even still, we observe an interesting synergistic performance improvement on the Lipids and Ion-Inserted Polymers subsets of the compositional test set whereby models trained on both \op{} and OMol25 meaningfully outperform models trained exclusively on either the polymer-only or molecule-only datasets. We also observe that these two categories exhibit additional fairly dramatic performance improvements with UMA-s-1p2, while for all other categories UMA-s-1p2 approaches but does not surpass the \op{}-only model. Indeed, UMA-s-1p2 is the only model to achieve chemical accuracy on Ion-Inserted configurations. Overall, these results indicate that polymer-specific data can be essential, synergistic, or at worst mildly helpful when incorporated alongside chemically diverse data, and that broader molecular data does not meaningfully degrade performance in polymer-specific contexts given a sufficiently expressive model.

\subsection{Evaluations}
\begin{table}[H]
\caption{Summary of model performances on \op{} evaluation tasks. On all three evaluation tasks, we report the mean absolute error in total energies (meV) and forces (meV/\angs). For the ion binding task, we report the MAE of the ion binding interaction (Ixn) energies and forces in addition to the MAE of the total energies and forces. All models achieve low errors on the distance scaling tasks, while the addition of \op{} data results in a noticeable performance improvement. Models trained only on \op{} achieve poor results on the ion binding task, likely due to the lack of training on significant amounts of ionic species. Nevertheless, jointly training on \op{} + OMol25 results in significantly improved performance over the OMol25 only and UMA-s-1p1 models.}
\centering
\resizebox{0.8\textwidth}{!}{%
\tablestyle{1.5pt}{1.05}
\begin{tabular}{y{50} ww ww ww ww}
% --- header row 1 (blank first cell) ---
\multicolumn{1}{c}{} 
& \multicolumn{2}{c}{
\ct[c3]{\it \parbox{1.35cm}{\centering Polymer\\Distance Scaling}}
}
& \multicolumn{2}{c}{
\ct[c4]{\it \parbox{1.35cm}{\centering Solvent\\Distance Scaling}}
}
& \multicolumn{4}{c}{\ct[c5]{\it Ion Binding}}
\\
% --- header row 2 (Model sits right above the data) ---
Model
& \cb[c3]{$\Delta$ Energy}{MAE [meV]$\downarrow$}
& \cb[c3]{$\Delta$ Forces}{MAE [meV/\angs]$\downarrow$} 
& \cb[c4]{$\Delta$ Energy}{MAE [meV]$\downarrow$}
& \cb[c4]{$\Delta$ Forces}{MAE [meV/\angs]$\downarrow$} 
& \cb[c5]{$\Delta$ Energy}{MAE [meV]$\downarrow$}
& \cb[c5]{$\Delta$ Forces}{MAE [meV/\angs]$\downarrow$}
& \cb[c5]{Ixn Energy}{MAE [meV]$\downarrow$}
& \cb[c5]{Ixn Forces}{MAE [meV/\angs]$\downarrow$}
\\
\hline
\addpadding
OMol25 Only  
& 1.20 & 0.58 
& 14.83 & 2.00 
& 64.07 & 6.43 & 133.7 & 6.01\\
\hline
\addpadding
UMA-s-1p1      
& 1.37 & 0.67 
& 9.92 & 2.28 
& 86.06 & 6.68 & 206.8 & 6.81\\
\midrule[1pt]
\op{} Only     
& 1.10 & 0.56 
& 9.00 & 2.20 
& 631.1 & 9.21 & 1695.8 & 12.16\\
\hline
\addpadding
\op{} + OMol25  
& 0.85 & 0.52 
& 10.13 & 1.89 
& 51.20 & 5.28 & 102.1 & 4.87\\
\hline
\addpadding
UMA-s-1p2  
& 0.73 & 0.48 
& 6.36 & 1.65 
& 33.28 & 4.73 & 73.1 & 4.35\\
\shline
\end{tabular}
}
\label{tab:omers_evals_results}
\end{table}

\textbf{Polymer distance scaling} evaluations show that all models are capable of capturing short-range interchain interactions with low energy MAE errors ( $\approx$ 1meV). Nevertheless, the model trained on \op{} + OMol25 achieves 29\% lower energy error than the model trained on OMol25 alone and 38\% lower than UMA-s-1p1. Further still, UMA-s-1p2 outperforms all other models, which may be due to model architecture improvements.

\textbf{Solvent distance scaling} provides a more nuanced view of the various models. Even without any explicit polymer training data, UMA-s-1p1 shows strong performance on this evaluation task, achieving an energy MAE comparable to the \op{} + OMol25 model. However, when comparing models of the same architecture, the \op{} + OMol25 model achieves 32\% lower energy MAE than the OMol25-only model.

\textbf{Ion Binding} shows the most significant improvement in model performance due to the inclusion of the \op{} data in training. Consistent with the results of Table \ref{tab:test_energy_errors}, we find a synergistic performance improvement on ion-polymer complexes where \op{} + OMol25 models significantly outperform models that have only seen polymers or molecules. A possible explanation for this effect is that these models are able to combine knowledge of local chemistry from small molecule datasets with interchain polymer interactions to capture both effects simultaneously. We note that models trained only on \op{} perform poorly on this task, which may be due to the lack of standalone charged species in \op{}. As with other tasks, UMA-s-1p2 achieves the best performance, likely due to a combination of its diverse training data and model architecture improvements over UMA-s-1p1.

\begin{table}[H]
\caption{Summary of model performances on OMol25 evaluation tasks. %Ixn energy/forces are the ligand-pocket interaction energy/forces of protein-ligand binding complex. IE and EA are the vertical ionization energy and electron affinity, respectively. Further 
Evaluation details are provided in the OMol25 paper \cite{levine2025openmolecules2025omol25}. Note that the first two models do not include \op{} in training, whereas the latter three models include \op{} data in training. Models trained on only OMol25 or \op{} + OMol25 perform comparably, demonstrating that the addition of polymer-specific data does not degrade performance in other chemical domains. A model trained solely on \op{} performs worse on or cannot reasonably perform tasks due to insufficient elemental/charge/spin coverage.}
\centering
\resizebox{\textwidth}{!}
{%
\tablestyle{1.5pt}{1.05}

\begin{tabular}{y{50} ww ww ww ww ww ww}
\multirow{2}{*}{\vspace{-2.9cm}Model} 
& \multicolumn{2}{c}{\ct[c3]{\it Ligand strain}} 
& \multicolumn{2}{c}{\ct[c4]{\it Conformers}} 
& \multicolumn{2}{c}{\ct[c5]{\it Protonation}}
& \multicolumn{2}{c}{\ct[c6]{\it Protein-Ligand}}
& \multicolumn{2}{c}{\ct[c3]{\it IE/EA}}
& \multicolumn{2}{c}{\ct[c4]{\it Spin gap}}\\
& \cb[c3]{Strain energy }{MAE [meV]$\downarrow$}
& \cb[c3]{RMSD}{min. [\angs~]$\downarrow$} 
& \cb[c4]{RMSD}{ensemble [\angs~]$\downarrow$} 
& \cb[c4]{$\Delta$ Energy}{MAE [meV]$\downarrow$}
& \cb[c5]{RMSD}{[\angs~]$\downarrow$} 
& \cb[c5]{$\Delta$ Energy}{MAE [meV]$\downarrow$}
& \cb[c6]{Ixn Energy}{MAE [meV]$\downarrow$} 
& \cb[c6]{Ixn Forces}{MAE [meV/\angs{}]$\downarrow$} 
& \cb[c3]{$\Delta$ Energy}{MAE [meV]$\downarrow$} 
& \cb[c3]{$\Delta$ Forces}{MAE [meV/\angs]$\downarrow$} 
& \cb[c4]{$\Delta$ Energy}{MAE [meV]$\downarrow$} 
& \cb[c4]{$\Delta$ Forces}{MAE [meV/\angs]$\downarrow$} 
\\
\hline
\addpadding
OMol25 Only& 4.6 & 0.25 & 0.04 & 5.25 & 0.06 & 23.4 & 209.4 & 4.82 & 202.8 & 54.9 & 314.0 & 56.4\\
\hline
\addpadding
UMA-s-1p1& 4.9 & 0.21 & 0.04 & 5.17 & 0.06 & 32.0 & 127.7 & 5.39 & 206.9 & 61.4 & 369.2 & 61.4\\
\midrule[1pt]
\addpadding
\op{} Only & 14.6 & 0.28 & 0.04 & 7.18 & {{---}} & {{---}} & {{---}} & {{---}} & {{---}} & {{---}} & {{---}} & {{---}}\\
\hline
\addpadding
\op{} + OMol25 & 5.2 & 0.22 & 0.04 & 5.30 & 0.06 & 29.2 & 190.0 & 4.74 & 200.9 & 55.6 & 297.5 & 57.4\\
\hline
\addpadding
UMA-s-1p2& 4.3 & 0.17 & 0.03 & 3.72 & 0.05 & 22.8 & 66.6 & 3.80 & 176.9 & 53.9 & 246.6 & 56.2\\
\shline
\end{tabular}
}
\label{tab:test_omol_ood_total}
\end{table}

As seen in Table \ref{tab:test_omol_ood_total}, the addition of \op{} data atop OMol25 data does not result in any significant degradation of performance on the OMol25 evaluations. The common level of theory of these two datasets and the complementary nature of the data accommodate broad performance with expressive models. However, a model trained only on \op{} does not contain comparable broad molecular chemical diversity or elemental/charge/spin distribution to OMol25 and so performs worse on or is inapplicable to many general molecular evaluation tasks.
\section{Conclusions and Future Directions}

%In this work, we presented the Open Polymers2026 dataset, a significant advance in data for training MLIPs for polymer applications. OPoly26 represents the first large scale open-source MLIP polymer dataset with a diverse range of polymer systems including linear copolymers, solvated polymers and reactive polymer structures.  We trained models that are broadly applicable to both small molecules and polymer systems without the need for system-specific fitting as is required for force field parametrization.  Models trained on this data show a marked improvement in polymer prediction tasks over baseline state-of-the-art molecular models, such as those trained on OMol25, without any significant performance degradation in molecular prediction tasks. We expect that the release of OPoly26 will facilitate future research into the generalization capabilities of small molecule models, while also highlighting current gaps in the chemical space of existing small molecule datasets.
Open Polymers 2026 (\op{}) is the first large-scale open-source DFT dataset for MLIP training with a diverse range of polymer systems including linear copolymers, solvated polymers and reactive polymer structures. We show that models trained on \op{} and OMol25 have small errors versus reference DFT for both small molecule and polymer system predictions without the need for system-specific tuning. Further, \op{} data markedly improves model polymer energy predictions compared with models trained only on broad molecular data (i.e. OMol25) without causing any significant performance degradation on molecular prediction tasks in other domains. Our analysis of how performance improvements from \op{} are distributed across polymer categories reveals the most significant improvements arising from reactive polymer configurations, in addition to smaller, but meaningful improvements across other polymer structures. We develop polymer-specific evaluations that demonstrate that models accurately capture near DFT-quality polymer-polymer and polymer-solvent interactions, and that the accuracy of polymer-ion interactions particularly benefit from highly expressive models that are jointly trained on both small molecule and polymer datasets. The release of \op{} therefore serves to fill in an important gap in the FF and MLIP landscape whereby systems containing both polymeric and molecular species, and which may involve chemical reactivity, can be accurately simulated without requiring any additional model tuning. In the future, we plan to examine how \op{} data impacts the capacity of models trained on \op{} to predict experimentally measured bulk polymer properties, leveraging the curated data in SimPoly. 

By enhancing MLIP performance across polymer domains, we envision that \op{} could enable accurate atomistic simulations of polymer dynamics and reactivity, facilitating computational study of polymer membranes for fuel cells and separations, polymer degradation pathways and upcycling, and polymer synthesis with minimal up-front burden for users. To foster community engagement and accelerate the development of generalizable polymer models, we release the \op{} dataset with an open-source CC-BY-4.0 license at \url{https://huggingface.co/facebook/OMol25} and will update the OMol25 leaderboard with polymer-centric tests and evaluations in the near future.

%Moving toward polymer models with strong generalization facilitates computational polymer design that can cover broad swaths of polymer chemical space. MLIPs are particularly promising due to their ability to handle bond-breaking and forming reactions, and therefore may open up the study of polymer degradation pathways, polymer upcycling, and polymer synthesis. Additional evaluations of MLIPs trained on \op{} will shed light on continuing challenges for polymer simulations.

%To help foster community engagement and accelerate the development of generalizable polymer models, we release the OPoly26 dataset with an open-source CC-BY-4.0 license at at \url{https://huggingface.co/facebook/OMol25}. We hope that this resource will encourage the community to conduct rigorous evaluations of model performance across diverse, real-world applications.
\section{Acknowledgments}

%%%E.R.A. acknowledges support from the Laboratory Directed Research and Development Program (LDRD) of Lawrence Livermore National Laboratory (LLNL), project numbers 24-SI-008 and 26-SI-004. E.R.A. also acknowledges the support of the Livermore Computing staff. N.L., J.D., S. S. M., and M.P.K. acknowledge support from the LDRD of LLNL, project number LDRD 23-ERD-030 and H.I.I. support from project number 24-ERD-027. This work was produced under the auspices of the U.S. Department of Energy by Lawrence Livermore National Laboratory under Contract DE-AC52-07NA27344.

We acknowledge support from the Laboratory Directed Research and Development Program of Lawrence Livermore National Laboratory, project numbers: E.R.A. 24-SI-008 and 26-SI-004; N.L., J.D., S. S. M., and M.P.K. 23-ERD-030; and H.I.I. 24-ERD-027. We acknowledge the support of the Livermore Computing staff. We also acknowledge useful discussions and feedback from Simon Pang, Anthony Varni, Sichi Li, Magi Yassa, and Johanna Schwartz. This work was produced under the auspices of the U.S. Department of Energy by Lawrence Livermore National Laboratory under Contract DE-AC52-07NA27344. 

L.C. acknowledges support by the U.S. Department of Energy, Office of Science, Office of Advanced Scientific Computing Research, Department of Energy Computational Science Graduate Fellowship under Award Number DE-SC0024386.

N.K. and S.M.B. acknowledge support from the Center for High Precision Patterning Science (CHiPPS), an Energy Frontier Research Center funded by the U.S. Department of Energy, Office of Science, Basic Energy Sciences at Lawrence Berkeley National Laboratory under Contract No. DE-AC02-34205CH11231. 

\section{Data availability}

The \op{} dataset is released under a CC-BY-4.0 license at \url{https://huggingface.co/facebook/OMol25} and code is available at \url{https://github.com/facebookresearch/fairchem}.

\clearpage
\newpage
\bibliographystyle{achemso}
\bibliography{paper}

@misc{levine2025openmolecules2025omol25,
      title={The Open Molecules 2025 (OMol25) Dataset, Evaluations, and Models}, 
      author={Daniel S. Levine and Muhammed Shuaibi and Evan Walter Clark Spotte-Smith and Michael G. Taylor and Muhammad R. Hasyim and Kyle Michel and Ilyes Batatia and Gábor Csányi and Misko Dzamba and Peter Eastman and Nathan C. Frey and Xiang Fu and Vahe Gharakhanyan and Aditi S. Krishnapriyan and Joshua A. Rackers and Sanjeev Raja and Ammar Rizvi and Andrew S. Rosen and Zachary Ulissi and Santiago Vargas and C. Lawrence Zitnick and Samuel M. Blau and Brandon M. Wood},
      year={2025},
      eprint={2505.08762},
      archivePrefix={arXiv},
      primaryClass={physics.chem-ph},
}

@article{hayashi_radonpy_2022,
	title = {{RadonPy}: automated physical property calculation using all-atom classical molecular dynamics simulations for polymer informatics},
	volume = {8},
	issn = {2057-3960},
	doi = {10.1038/s41524-022-00906-4},
	number = {1},
	journal = {npj Computational Materials},
	author = {Hayashi, Yoshihiro and Shiomi, Junichiro and Morikawa, Junko and Yoshida, Ryo},
	month = nov,
	year = {2022},
	pages = {222},
}

@article{lohmann_PFAS_2020,
	title = {Are {Fluoropolymers} {Really} of {Low} {Concern} for {Human} and {Environmental} {Health} and {Separate} from {Other} {PFAS}?},
	volume = {54},
	issn = {0013-936X},
	doi = {10.1021/acs.est.0c03244},
	number = {20},
	urldate = {2025-09-09},
	journal = {Environmental Science \& Technology},
	author = {Lohmann, Rainer and Cousins, Ian T. and DeWitt, Jamie C. and Glüge, Juliane and Goldenman, Gretta and Herzke, Dorte and Lindstrom, Andrew B. and Miller, Mark F. and Ng, Carla A. and Patton, Sharyle and Scheringer, Martin and Trier, Xenia and Wang, Zhanyun},
	month = oct,
	year = {2020},
	pages = {12820--12828},
}

@article{ebiomedicine_forever_2023,
	title = {Forever chemicals: the persistent effects of perfluoroalkyl and polyfluoroalkyl substances on human health},
	volume = {95},
	issn = {2352-3964},
	shorttitle = {Forever chemicals},
	doi = {10.1016/j.ebiom.2023.104806},
	language = {English},
	urldate = {2025-09-09},
	journal = {eBioMedicine},
	author = {eBioMedicine},
	month = sep,
	year = {2023},
	pmid = {37714648},
}

@book{rasmussen_conjugated_2023,
	address = {Washington, DC, USA},
	title = {Conjugated {Polymers}: {Synthesis} \& {Design}},
	publisher = {American Chemical Society},
	author = {Rasmussen, Seth C. and Gilman, Spencer J. and Wilcox, Wyatt D.},
	year = {2023},
	doi = {10.1021/acsinfocus.7e7026},
}

@article{xu_new_2021,
	title = {A {New} {Conjugated} {Polymer} that {Enables} the {Integration} of {Photovoltaic} and {Light}-{Emitting} {Functions} in {One} {Device}},
	volume = {33},
	copyright = {© 2021 Wiley-VCH GmbH},
	issn = {1521-4095},
	doi = {10.1002/adma.202101090},
	language = {en},
	number = {22},
	urldate = {2025-09-09},
	journal = {Advanced Materials},
	author = {Xu, Ye and Cui, Yong and Yao, Huifeng and Zhang, Tao and Zhang, Jianqi and Ma, Lijiao and Wang, Jingwen and Wei, Zhixiang and Hou, Jianhui},
	year = {2021},
	pages = {2101090},
}

@article{zhang_thiadiazole-based_2021,
	title = {A {Thiadiazole}-{Based} {Conjugated} {Polymer} with {Ultradeep} {HOMO} {Level} and {Strong} {Electroluminescence} {Enables} 18.6\% {Efficiency} in {Organic} {Solar} {Cell}},
	volume = {11},
	copyright = {© 2021 Wiley-VCH GmbH},
	issn = {1614-6840},
	doi = {10.1002/aenm.202101705},
	language = {en},
	number = {35},
	urldate = {2025-09-09},
	journal = {Advanced Energy Materials},
	author = {Zhang, Tao and An, Cunbin and Bi, Pengqing and Lv, Qianglong and Qin, Jinzhao and Hong, Ling and Cui, Yong and Zhang, Shaoqing and Hou, Jianhui},
	year = {2021},
	pages = {2101705},
}

@article{bai_accelerated_2019,
	title = {Accelerated {Discovery} of {Organic} {Polymer} {Photocatalysts} for {Hydrogen} {Evolution} from {Water} through the {Integration} of {Experiment} and {Theory}},
	volume = {141},
	issn = {0002-7863},
	doi = {10.1021/jacs.9b03591},
	number = {22},
	urldate = {2025-09-09},
	journal = {Journal of the American Chemical Society},
	author = {Bai, Yang and Wilbraham, Liam and Slater, Benjamin J. and Zwijnenburg, Martijn A. and Sprick, Reiner Sebastian and Cooper, Andrew I.},
	month = jun,
	year = {2019},
	pages = {9063--9071},
}

@article{aldeghi_graph_2022,
	title = {A graph representation of molecular ensembles for polymer property prediction},
	volume = {13},
	issn = {2041-6539},
	url = {https://pubs.rsc.org/en/content/articlelanding/2022/sc/d2sc02839e},
	doi = {10.1039/D2SC02839E},
	language = {en},
	number = {35},
	urldate = {2025-09-09},
	journal = {Chemical Science},
	author = {Aldeghi, Matteo and Coley, Connor W.},
	month = sep,
	year = {2022},
}

@article{kim_OMG_2023,
	title = {Open {Macromolecular} {Genome}: {Generative} {Design} of {Synthetically} {Accessible} {Polymers}},
	volume = {3},
	shorttitle = {Open {Macromolecular} {Genome}},
	url = {https://doi.org/10.1021/acspolymersau.3c00003},
	doi = {10.1021/acspolymersau.3c00003},
	number = {4},
	urldate = {2025-09-09},
	journal = {ACS Polymers Au},
	author = {Kim, Seonghwan and Schroeder, Charles M. and Jackson, Nicholas E.},
	month = aug,
	year = {2023},
	pages = {318--330},
}

@misc{rdkit,
  title        = {RDKit: Open-source cheminformatics},
  author       = {{RDKit community}},
  year         = {2025},
  howpublished = {\url{https://www.rdkit.org}},
}

@article{xie_accelerating_2022,
	title = {Accelerating amorphous polymer electrolyte screening by learning to reduce errors in molecular dynamics simulated properties},
	volume = {13},
	copyright = {2022 The Author(s)},
	issn = {2041-1723},
	url = {https://www.nature.com/articles/s41467-022-30994-1},
	doi = {10.1038/s41467-022-30994-1},
	language = {en},
	number = {1},
	urldate = {2025-11-04},
	journal = {Nature Communications},
	author = {Xie, Tian and France-Lanord, Arthur and Wang, Yanming and Lopez, Jeffrey and Stolberg, Michael A. and Hill, Megan and Leverick, Graham Michael and Gomez-Bombarelli, Rafael and Johnson, Jeremiah A. and Shao-Horn, Yang and Grossman, Jeffrey C.},
	month = jun,
	year = {2022},
	pages = {3415},
}

@article{song_reflection_2023,
	title = {A reflection on polymer electrolytes for solid-state lithium metal batteries},
	volume = {14},
	copyright = {2023 The Author(s)},
	issn = {2041-1723},
	url = {https://www.nature.com/articles/s41467-023-40609-y},
	doi = {10.1038/s41467-023-40609-y},
	language = {en},
	number = {1},
	urldate = {2025-11-04},
	journal = {Nature Communications},
	author = {Song, Ziyu and Chen, Fangfang and Martinez-Ibañez, Maria and Feng, Wenfang and Forsyth, Maria and Zhou, Zhibin and Armand, Michel and Zhang, Heng},
	month = aug,
	year = {2023},
	pages = {4884},
}

@article{matsumura_generator_2025,
	title = {Generator of {Neural} {Network} {Potential} for {Molecular} {Dynamics}: {Constructing} {Robust} and {Accurate} {Potentials} with {Active} {Learning} for {Nanosecond}-{Scale} {Simulations}},
	volume = {21},
	issn = {1549-9618},
	shorttitle = {Generator of {Neural} {Network} {Potential} for {Molecular} {Dynamics}},
	url = {https://doi.org/10.1021/acs.jctc.4c01613},
	doi = {10.1021/acs.jctc.4c01613},
	number = {8},
	urldate = {2025-11-04},
	journal = {Journal of Chemical Theory and Computation},
	author = {Matsumura, Naoki and Yoshimoto, Yuta and Yamazaki, Tamio and Amano, Tomohito and Noda, Tomoyuki and Ebata, Naoki and Kasano, Takatoshi and Sakai, Yasufumi},
	month = apr,
	year = {2025},
	pages = {3832--3846},
}

@book{Ramsundar-et-al-2019,
    title={Deep Learning for the Life Sciences},
    author={Bharath Ramsundar and Peter Eastman and Patrick Walters and Vijay Pande and Karl Leswing and Zhenqin Wu},
    year={2019}
}

@article{mermin_thermal_1965,
	title = {Thermal {Properties} of the {Inhomogeneous} {Electron} {Gas}},
	volume = {137},
	url = {https://link.aps.org/doi/10.1103/PhysRev.137.A1441},
	doi = {10.1103/PhysRev.137.A1441},
	number = {5A},
	urldate = {2025-12-23},
	journal = {Physical Review},
	author = {Mermin, N. David},
	month = mar,
	year = {1965},
	pages = {A1441--A1443},
}

@article{gaus_parameterization_2014,
	title = {Parameterization of {DFTB3}/{3OB} for {Sulfur} and {Phosphorus} for {Chemical} and {Biological} {Applications}},
	volume = {10},
	issn = {1549-9618},
	url = {https://doi.org/10.1021/ct401002w},
	doi = {10.1021/ct401002w},
	number = {4},
	urldate = {2025-12-23},
	journal = {Journal of Chemical Theory and Computation},
	author = {Gaus, Michael and Lu, Xiya and Elstner, Marcus and Cui, Qiang},
	month = apr,
	year = {2014},
	pages = {1518--1537},
}

@article{hourahine_dftb_2020,
	title = {{DFTB}+, a software package for efficient approximate density functional theory based atomistic simulations},
	volume = {152},
	issn = {0021-9606},
	url = {https://doi.org/10.1063/1.5143190},
	doi = {10.1063/1.5143190},
	number = {12},
	urldate = {2025-12-23},
	journal = {The Journal of Chemical Physics},
	author = {Hourahine, B. and Aradi, B. and Blum, V. and Bonafé, F. and Buccheri, A. and Camacho, C. and Cevallos, C. and Deshaye, M. Y. and Dumitrică, T. and Dominguez, A. and Ehlert, S. and Elstner, M. and van der Heide, T. and Hermann, J. and Irle, S. and Kranz, J. J. and Köhler, C. and Kowalczyk, T. and Kubař, T. and Lee, I. S. and Lutsker, V. and Maurer, R. J. and Min, S. K. and Mitchell, I. and Negre, C. and Niehaus, T. A. and Niklasson, A. M. N. and Page, A. J. and Pecchia, A. and Penazzi, G. and Persson, M. P. and Řezáč, J. and Sánchez, C. G. and Sternberg, M. and Stöhr, M. and Stuckenberg, F. and Tkatchenko, A. and Yu, V. W.-z. and Frauenheim, T.},
	month = mar,
	year = {2020},
	pages = {124101},
}

@article{gaus_dftb3_2011,
	title = {{DFTB3}: {Extension} of the {Self}-{Consistent}-{Charge} {Density}-{Functional} {Tight}-{Binding} {Method} ({SCC}-{DFTB})},
	volume = {7},
	issn = {1549-9618},
	shorttitle = {{DFTB3}},
	url = {https://doi.org/10.1021/ct100684s},
	doi = {10.1021/ct100684s},
	number = {4},
	urldate = {2025-12-23},
	journal = {Journal of Chemical Theory and Computation},
	author = {Gaus, Michael and Cui, Qiang and Elstner, Marcus},
	month = apr,
	year = {2011},
	pages = {931--948},
}

@article{bayly_well-behaved_1993,
	title = {A well-behaved electrostatic potential based method using charge restraints for deriving atomic charges: the {RESP} model},
	volume = {97},
	issn = {0022-3654},
	shorttitle = {A well-behaved electrostatic potential based method using charge restraints for deriving atomic charges},
	url = {https://doi.org/10.1021/j100142a004},
	doi = {10.1021/j100142a004},
	number = {40},
	urldate = {2025-12-23},
	journal = {The Journal of Physical Chemistry},
	author = {Bayly, Christopher I. and Cieplak, Piotr and Cornell, Wendy and Kollman, Peter A.},
	month = oct,
	year = {1993},
	pages = {10269--10280},
}

@article{melenkevitz_density_1991,
	title = {Density functional theory of lamellar ordering in diblock copolymers},
	volume = {24},
	issn = {0024-9297},
	url = {https://doi.org/10.1021/ma00014a038},
	doi = {10.1021/ma00014a038},
	number = {14},
	urldate = {2025-12-22},
	journal = {{Macromolecules}},
	author = {Melenkevitz, J. and Muthukumar, M.},
	month = jul,
	year = {1991},
	pages = {4199--4205},
}

@article{lei2023self,
  title={A self-improvable Polymer Discovery Framework Based on Conditional Generative Model},
  author={Lei, Xiangyun and Ye, Weike and Yang, Zhenze and Schweigert, Daniel and Kwon, Ha-Kyung and Khajeh, Arash},
  journal={arXiv preprint arXiv:2312.04013},
  year={2023}
}

@article{gonzalez_cortes_effect_2023,
	title = {Effect of intermolecular interactions on the glass transition temperature of chemically modified alternating polyketones},
	volume = {34},
	issn = {2468-5194},
	url = {https://www.sciencedirect.com/science/article/pii/S2468519423003981},
	doi = {10.1016/j.mtchem.2023.101771},
	urldate = {2026-03-01},
	journal = {Materials Today Chemistry},
	author = {González Cortes, Pablo and Araya-Hermosilla, Rodrigo and Wrighton-Araneda, Kerry and Cortés-Arriagada, Diego and Picchioni, Francesco and Yan, Feng and Rudolf, Petra and Bose, Ranjita K. and Quero, Franck},
	month = dec,
	year = {2023},
	pages = {101771},
}

@article{beljonne_interchain_2000,
	title = {Interchain interactions in conjugated materials: {The} exciton model versus the supermolecular approach},
	volume = {112},
	issn = {0021-9606},
	shorttitle = {Interchain interactions in conjugated materials},
	url = {https://doi.org/10.1063/1.481031},
	doi = {10.1063/1.481031},
	number = {10},
	urldate = {2026-03-01},
	journal = {The Journal of Chemical Physics},
	author = {Beljonne, D. and Cornil, J. and Silbey, R. and Millié, P. and Brédas, J. L.},
	month = mar,
	year = {2000},
	pages = {4749--4758},
}

@article{ishizaka_minor_2024,
	title = {A minor difference in the hydrogen-bonding group structure has a major impact on the mechanical properties of polymers},
	volume = {15},
	issn = {1759-9962},
	url = {https://pubs.rsc.org/en/content/articlelanding/2024/py/d4py00580e},
	doi = {10.1039/D4PY00580E},
	language = {en},
	number = {39},
	urldate = {2026-03-01},
	journal = {Polymer Chemistry},
	publisher = {The Royal Society of Chemistry},
	author = {Ishizaka, Shogo and Nakagawa, Shintaro and Yoshie, Naoko},
	month = oct,
	year = {2024},
	pages = {3967--3976},
}

@article{Larsen_2017,
	author = {Larsen, Ask Hjorth and Mortensen, Jens Jørgen and Blomqvist, Jakob and Castelli, Ivano E. and Christensen, Rune and Du{\l}ak, Marcin and Friis, Jesper and Groves, Michael N. and Hammer, Bjørk and Hargus, Cory and Hermes, Eric D. and Jennings, Paul C. and Jensen, Peter Bjerre and Kermode, James and Kitchin, John R. and Kolsbjerg, Esben Leonhard and Kubal, Joseph and Kaasbjerg, Kristen and Lysgaard, Steen and Maronsson, Jón Bergmann and Maxson, Tristan and Olsen, Thomas and Pastewka, Lars and Peterson, Andrew and Rostgaard, Carsten and Schiøtz, Jakob and Schütt, Ole and Strange, Mikkel and Thygesen, Kristian S. and Vegge, Tejs and Vilhelmsen, Lasse and Walter, Michael and Zeng, Zhenhua and Jacobsen, Karsten Wedel},
	title = {The Atomic Simulation Environment—A Python library for working with atoms},
	journal = {Journal of Physics: Condensed Matter},
	volume = {29},
	number = {27},
	pages = {273002},
	year = {2017},
	doi = {10.1088/1361-648X/aa680e}
}

@misc{batatia_mace_2023,
	title = {{MACE}: {Higher} {Order} {Equivariant} {Message} {Passing} {Neural} {Networks} for {Fast} and {Accurate} {Force} {Fields}},
	shorttitle = {{MACE}},
	doi = {10.48550/arXiv.2206.07697},
	urldate = {2025-12-24},
	publisher = {arXiv},
	author = {Batatia, Ilyes and Kovács, Dávid Péter and Simm, Gregor N. C. and Ortner, Christoph and Csányi, Gábor},
	month = jan,
	year = {2023},
	note = {arXiv:2206.07697 [stat]},
}

@article{vattulainen_lipid_2011,
	title = {Lipid {Simulations}: {A} {Perspective} on {Lipids} in {Action}},
	volume = {3},
	issn = {1943-0264},
	shorttitle = {Lipid {Simulations}},
	doi = {10.1101/cshperspect.a004655},
	number = {4},
	urldate = {2025-12-24},
	journal = {Cold Spring Harbor Perspectives in Biology},
	author = {Vattulainen, Ilpo and Rog, Tomasz},
	month = apr,
	year = {2011},
	pmid = {21441592},
	pmcid = {PMC3062216},
	pages = {a004655},
}

@misc{chiang_mlip_2025,
	title = {{MLIP} {Arena}: {Advancing} {Fairness} and {Transparency} in {Machine} {Learning} {Interatomic} {Potentials} via an {Open}, {Accessible} {Benchmark} {Platform}},
	shorttitle = {{MLIP} {Arena}},
	doi = {10.48550/arXiv.2509.20630},
	urldate = {2025-12-24},
	publisher = {arXiv},
	author = {Chiang, Yuan and Kreiman, Tobias and Zhang, Christine and Kuner, Matthew C. and Weaver, Elizabeth and Amin, Ishan and Park, Hyunsoo and Lim, Yunsung and Kim, Jihan and Chrzan, Daryl and Walsh, Aron and Blau, Samuel M. and Asta, Mark and Krishnapriyan, Aditi S.},
	month = nov,
	year = {2025},
	note = {arXiv:2509.20630 [physics]},
}

@misc{rhodes_orb-v3_2025,
	title = {Orb-v3: atomistic simulation at scale},
	shorttitle = {Orb-v3},
	doi = {10.48550/arXiv.2504.06231},
	urldate = {2025-12-24},
	publisher = {arXiv},
	author = {Rhodes, Benjamin and Vandenhaute, Sander and Šimkus, Vaidotas and Gin, James and Godwin, Jonathan and Duignan, Tim and Neumann, Mark},
	month = apr,
	year = {2025},
	note = {arXiv:2504.06231 [cond-mat]},
}

@article{mayo_dreiding_1990,
	title = {{DREIDING}:  a generic force field for molecular simulations},
	volume = {94},
	issn = {0022-3654},
	shorttitle = {{DREIDING}},
	doi = {10.1021/j100389a010},
	number = {26},
	urldate = {2025-12-24},
	journal = {The Journal of Physical Chemistry},
	author = {Mayo, Stephen L. and Olafson, Barry D. and Goddard, William A.},
	month = dec,
	year = {1990},
	note = {Publisher: American Chemical Society},
	pages = {8897--8909},
}

@article{kaminski_evaluation_2001,
	title = {Evaluation and {Reparametrization} of the {OPLS}-{AA} {Force} {Field} for {Proteins} via {Comparison} with {Accurate} {Quantum} {Chemical} {Calculations} on {Peptides}},
	volume = {105},
	issn = {1520-6106},
	url = {https://doi.org/10.1021/jp003919d},
	doi = {10.1021/jp003919d},
	number = {28},
	urldate = {2025-12-24},
	journal = {The Journal of Physical Chemistry B},
	author = {Kaminski, George A. and Friesner, Richard A. and Tirado-Rives, Julian and Jorgensen, William L.},
	month = jul,
	year = {2001},
	note = {Publisher: American Chemical Society},
	pages = {6474--6487},
}

@article{yang2023novo,
  title={De novo design of polymer electrolytes with high conductivity using gpt-based and diffusion-based generative models},
  author={Yang, Zhenze and Ye, Weike and Lei, Xiangyun and Schweigert, Daniel and Kwon, Ha-Kyung and Khajeh, Arash},
  journal={arXiv preprint arXiv:2312.06470},
  year={2023}
}

@article{jorgensen_development_1996,
	title = {Development and {Testing} of the {OPLS} {All}-{Atom} {Force} {Field} on {Conformational} {Energetics} and {Properties} of {Organic} {Liquids}},
	volume = {118},
	issn = {0002-7863},
	doi = {10.1021/ja9621760},
	number = {45},
	urldate = {2025-12-24},
	journal = {Journal of the American Chemical Society},
	author = {Jorgensen, William L. and Maxwell, David S. and Tirado-Rives, Julian},
	month = nov,
	year = {1996},
	pages = {11225--11236},
}

@article{bishop_molecular_1979,
	title = {Molecular dynamics of polymeric systems},
	volume = {70},
	issn = {0021-9606},
	doi = {10.1063/1.437567},
	number = {3},
	urldate = {2025-12-24},
	journal = {The Journal of Chemical Physics},
	author = {Bishop, Marvin and Kalos, M. H. and Frisch, H. L.},
	month = feb,
	year = {1979},
	pages = {1299--1304},
}

@article{gooneie_review_2017,
	title = {A {Review} of {Multiscale} {Computational} {Methods} in {Polymeric} {Materials}},
	volume = {9},
	copyright = {http://creativecommons.org/licenses/by/3.0/},
	issn = {2073-4360},
	doi = {10.3390/polym9010016},
	language = {en},
	number = {1},
	urldate = {2025-12-24},
	journal = {Polymers},
	author = {Gooneie, Ali and Schuschnigg, Stephan and Holzer, Clemens},
	month = jan,
	year = {2017},
	pages = {16},
}

@article{gartner_modeling_2019,
	title = {Modeling and {Simulations} of {Polymers}: {A} {Roadmap}},
	volume = {52},
	issn = {0024-9297},
	shorttitle = {Modeling and {Simulations} of {Polymers}},
	url = {https://doi.org/10.1021/acs.macromol.8b01836},
	doi = {10.1021/acs.macromol.8b01836},
	number = {3},
	urldate = {2025-12-22},
	journal = {Macromolecules},
	author = {Gartner, Thomas E. III and Jayaraman, Arthi},
	month = feb,
	year = {2019},
	pages = {755--786},
}

@article{chen_covalent_2024,
	title = {Covalent adaptable polymer networks with {CO2}-facilitated recyclability},
	volume = {15},
	copyright = {2024 The Author(s)},
	issn = {2041-1723},
	url = {https://www.nature.com/articles/s41467-024-50738-7},
	doi = {10.1038/s41467-024-50738-7},
	language = {en},
	number = {1},
	urldate = {2025-12-22},
	journal = {Nature Communications},
	author = {Chen, Jiayao and Li, Lin and Luo, Jiancheng and Meng, Lingyao and Zhao, Xiao and Song, Shenghan and Demchuk, Zoriana and Li, Pei and He, Yi and Sokolov, Alexei P. and Cao, Peng-Fei},
	month = aug,
	year = {2024},
	pages = {6605},
}

@article{xie_hydrogen_2021,
	title = {Hydrogen {Bonding} in {Self}-{Healing} {Elastomers}},
	volume = {6},
	url = {https://doi.org/10.1021/acsomega.1c00462},
	doi = {10.1021/acsomega.1c00462},
	number = {14},
	urldate = {2025-12-22},
	journal = {ACS Omega},
	author = {Xie, Zhulu and Hu, Ben-Lin and Li, Run-Wei and Zhang, Qichun},
	month = apr,
	year = {2021},
	pages = {9319--9333},
}

@article{song_high-performance_2020,
	title = {High-{Performance} {Polymeric} {Materials} through {Hydrogen}-{Bond} {Cross}-{Linking}},
	volume = {32},
	copyright = {© 2019 WILEY-VCH Verlag GmbH \& Co. KGaA, Weinheim},
	issn = {1521-4095},
	doi = {10.1002/adma.201901244},
	language = {en},
	number = {18},
	urldate = {2025-12-22},
	journal = {Advanced Materials},
	author = {Song, Pingan and Wang, Hao},
	year = {2020},
	pages = {1901244},
}

@article{li_potential_2023,
	title = {Potential {Health} {Impact} of {Microplastics}: {A} {Review} of {Environmental} {Distribution}, {Human} {Exposure}, and {Toxic} {Effects}},
	volume = {1},
	shorttitle = {Potential {Health} {Impact} of {Microplastics}},
	url = {https://doi.org/10.1021/envhealth.3c00052},
	doi = {10.1021/envhealth.3c00052},
	number = {4},
	urldate = {2025-12-22},
	journal = {Environment \& Health},
	author = {Li, Yue and Tao, Le and Wang, Qiong and Wang, Fengbang and Li, Gang and Song, Maoyong},
	month = oct,
	year = {2023},
	pages = {249--257},
}

@article{jambeck_plastic_2015,
	title = {Plastic waste inputs from land into the ocean},
	volume = {347},
	url = {https://www.science.org/doi/10.1126/science.1260352},
	doi = {10.1126/science.1260352},
	number = {6223},
	urldate = {2025-12-22},
	journal = {Science},
	author = {Jambeck, Jenna R. and Geyer, Roland and Wilcox, Chris and Siegler, Theodore R. and Perryman, Miriam and Andrady, Anthony and Narayan, Ramani and Law, Kara Lavender},
	month = feb,
	year = {2015},
	pages = {768--771},
}

@article{sun_ab_1994,
	title = {An ab {Initio} {CFF93} {All}-{Atom} {Force} {Field} for {Polycarbonates}},
	volume = {116},
	issn = {0002-7863},
	url = {https://doi.org/10.1021/ja00086a030},
	doi = {10.1021/ja00086a030},
	number = {7},
	urldate = {2025-11-04},
	journal = {Journal of the American Chemical Society},
	author = {Sun, Huai and Mumby, Stephen J. and Maple, Jon R. and Hagler, Arnold T.},
	month = apr,
	year = {1994},
	pages = {2978--2987},
}

@article{hansen_universality_1969,
	title = {The {Universality} of the {Solubility} {Parameter}},
	volume = {8},
	issn = {0091-1968},
	url = {https://doi.org/10.1021/i360029a002},
	doi = {10.1021/i360029a002},
	number = {1},
	urldate = {2026-03-01},
	journal = {Product R\&D},
	publisher = {American Chemical Society},
	author = {Hansen, Charles M.},
	month = mar,
	year = {1969},
	pages = {2--11},
}

@article{da_luz_polyionic_2022,
	title = {Poly(ionic liquid)s-based polyurethane blends: effect of polyols structure and {ILs} counter cations in {CO2} sorption performance of {PILs} physical blends},
	volume = {79},
	issn = {1436-2449},
	shorttitle = {Poly(ionic liquid)s-based polyurethane blends},
	url = {https://doi.org/10.1007/s00289-021-03799-3},
	doi = {10.1007/s00289-021-03799-3},
	language = {en},
	number = {8},
	urldate = {2026-03-01},
	journal = {Polymer Bulletin},
	author = {da Luz, Murilo and Dias, Guilherme and Zimmer, Henrique and Bernard, Franciele L. and do Nascimento, Jailton F. and Einloft, Sandra},
	month = aug,
	year = {2022},
	pages = {6123--6139},
}

@article{allam_molecular_2025,
	title = {Molecular {Insights} into {Lithium}-{Ion} {Coordination} and {Morphology} in {Carbonate} {Polymer} {Electrolytes}},
	volume = {37},
	issn = {0897-4756},
	url = {https://doi.org/10.1021/acs.chemmater.5c01016},
	doi = {10.1021/acs.chemmater.5c01016},
	number = {17},
	urldate = {2026-03-01},
	journal = {Chemistry of Materials},
	publisher = {American Chemical Society},
	author = {Allam, Omar and Jang, Seung Soon},
	month = sep,
	year = {2025},
	pages = {6574--6584},
}

@article{kozuch_predicting_2016,
	title = {Predicting the {Flory}-{Huggins} χ {Parameter} for {Polymers} with {Stiffness} {Mismatch} from {Molecular} {Dynamics} {Simulations}},
	volume = {8},
	issn = {2073-4360},
	doi = {10.3390/polym8060241},
	language = {eng},
	number = {6},
	journal = {Polymers},
	author = {Kozuch, Daniel J. and Zhang, Wenlin and Milner, Scott T.},
	month = jun,
	year = {2016},
	pages = {241},
}

@article{flory_thermodynamics_1941,
	title = {Thermodynamics of {High} {Polymer} {Solutions}},
	volume = {9},
	issn = {0021-9606},
	url = {https://doi.org/10.1063/1.1750971},
	doi = {10.1063/1.1750971},
	number = {8},
	urldate = {2025-11-15},
	journal = {The Journal of Chemical Physics},
	author = {Flory, Paul J.},
	month = aug,
	year = {1941},
	pages = {660},
}

@article{ethier_predicting_2024,
	title = {Predicting polymer solubility from phase diagrams to compatibility: a perspective on challenges and opportunities},
	volume = {20},
	issn = {1744-6848},
	shorttitle = {Predicting polymer solubility from phase diagrams to compatibility},
	url = {https://pubs.rsc.org/en/content/articlelanding/2024/sm/d4sm00590b},
	doi = {10.1039/D4SM00590B},
	language = {en},
	number = {29},
	urldate = {2025-11-15},
	journal = {Soft Matter},
	author = {Ethier, Jeffrey and Antoniuk, Evan R. and Brettmann, Blair},
	month = jul,
	year = {2024},
	pages = {5652--5669},
}

@article{huggins_theory_1942,
	title = {Theory of {Solutions} of {High} {Polymers1}},
	volume = {64},
	issn = {0002-7863},
	url = {https://doi.org/10.1021/ja01259a068},
	doi = {10.1021/ja01259a068},
	number = {7},
	urldate = {2025-11-15},
	journal = {Journal of the American Chemical Society},
	author = {Huggins, Maurice L.},
	month = jul,
	year = {1942},
	pages = {1712--1719},
}

@misc{fu_learning_2025,
	title = {Learning {Smooth} and {Expressive} {Interatomic} {Potentials} for {Physical} {Property} {Prediction}},
	url = {http://arxiv.org/abs/2502.12147},
	doi = {10.48550/arXiv.2502.12147},
	urldate = {2025-11-12},
	publisher = {arXiv},
	author = {Fu, Xiang and Wood, Brandon M. and Barroso-Luque, Luis and Levine, Daniel S. and Gao, Meng and Dzamba, Misko and Zitnick, C. Lawrence},
	month = apr,
	year = {2025},
	note = {arXiv:2502.12147 [physics]},
}

@article{sun_compass_1998,
	title = {{COMPASS}: {An} ab {Initio} {Force}-{Field} {Optimized} for {Condensed}-{Phase} {ApplicationsOverview} with {Details} on {Alkane} and {Benzene} {Compounds}},
	volume = {102},
	issn = {1520-6106},
	shorttitle = {{COMPASS}},
	url = {https://doi.org/10.1021/jp980939v},
	doi = {10.1021/jp980939v},
	number = {38},
	urldate = {2025-11-10},
	journal = {The Journal of Physical Chemistry B},
	author = {Sun, H.},
	month = sep,
	year = {1998},
	pages = {7338--7364},
}

@article{levine_large_2023,
	title = {Large {Computational} {Survey} of {Intrinsic} {Reactivity} of {Aromatic} {Carbon} {Atoms} with {Respect} to a {Model} {Aldehyde} {Oxidase}},
	volume = {19},
	issn = {1549-9618},
	url = {https://doi.org/10.1021/acs.jctc.3c00913},
	doi = {10.1021/acs.jctc.3c00913},
	number = {24},
	urldate = {2024-03-29},
	journal = {Journal of Chemical Theory and Computation},
	author = {Levine, Daniel S. and Jacobson, Leif D. and Bochevarov, Art D.},
	month = dec,
	year = {2023},
	note = {Publisher: American Chemical Society},
	pages = {9302--9317}
}

@inproceedings{otsuka_polyinfo_2011,
	title = {{PoLyInfo}: {Polymer} {Database} for {Polymeric} {Materials} {Design}},
	shorttitle = {{PoLyInfo}},
	doi = {10.1109/EIDWT.2011.13},
	urldate = {2025-11-01},
	booktitle = {2011 {International} {Conference} on {Emerging} {Intelligent} {Data} and {Web} {Technologies}},
	author = {Otsuka, Shingo and Kuwajima, Isao and Hosoya, Junko and Xu, Yibin and Yamazaki, Masayoshi},
	month = sep,
	year = {2011},
	pages = {22--29},
}

@book{bicerano_prediction_2002,
	address = {Boca Raton},
	edition = {3},
	title = {Prediction of {Polymer} {Properties}},
	isbn = {978-0-429-22228-3},
	publisher = {CRC Press},
	author = {Bicerano, Jozef},
	month = jul,
	year = {2002},
	doi = {10.1201/9780203910115},
}

@article{price_tip3p_modified_2004,
	title = {A modified {TIP3P} water potential for simulation with {Ewald} summation},
	volume = {121},
	issn = {0021-9606},
	url = {https://doi.org/10.1063/1.1808117},
	doi = {10.1063/1.1808117},
	number = {20},
	urldate = {2025-10-31},
	journal = {The Journal of Chemical Physics},
	author = {Price, Daniel J. and Brooks, III, Charles L.},
	month = nov,
	year = {2004},
	pages = {10096--10103},
}

@article{trag_improved_gaff2_2019,
	title = {Improved {GAFF2} parameters for fluorinated alkanes and mixed hydro- and fluorocarbons},
	volume = {25},
	issn = {0948-5023},
	doi = {10.1007/s00894-018-3911-5},
	number = {2},
	journal = {Journal of Molecular Modeling},
	author = {Träg, Johannes and Zahn, Dirk},
	month = jan,
	year = {2019},
	pages = {39},
}

@article{mardirossian_b97m-v_2016,
	title = {ω{B97M}-{V}: {A} combinatorially optimized, range-separated hybrid, meta-{GGA} density functional with {VV10} nonlocal correlation},
	volume = {144},
	issn = {0021-9606},
	shorttitle = {ω{B97M}-{V}},
	doi = {10.1063/1.4952647},
	number = {21},
	urldate = {2025-11-07},
	journal = {The Journal of Chemical Physics},
	author = {Mardirossian, Narbe and Head-Gordon, Martin},
	month = jun,
	year = {2016},
	pages = {214110},
}

@article{weigend_balanced_2005,
	title = {Balanced basis sets of split valence, triple zeta valence and quadruple zeta valence quality for {H} to {Rn}: {Design} and assessment of accuracy},
	volume = {7},
	issn = {1463-9084},
	shorttitle = {Balanced basis sets of split valence, triple zeta valence and quadruple zeta valence quality for {H} to {Rn}},
	url = {https://pubs.rsc.org/en/content/articlelanding/2005/cp/b508541a},
	doi = {10.1039/B508541A},
	language = {en},
	number = {18},
	urldate = {2025-11-07},
	journal = {Physical Chemistry Chemical Physics},
	author = {Weigend, Florian and Ahlrichs, Reinhart},
	month = aug,
	year = {2005},
	pages = {3297--3305},
}

@article{kroonblawd_polymer_2022,
	title = {Polymer degradation through chemical change: a quantum-based test of inferred reactions in irradiated polydimethylsiloxane},
	shorttitle = {Polymer degradation through chemical change},
	url = {https://pubs.rsc.org/en/content/articlehtml/2022/cp/d1cp05647f},
	doi = {10.1039/D1CP05647F},
	language = {en},
	urldate = {2025-11-08},
	author = {Kroonblawd, Matthew and Goldman, Nir and Maiti, Amitesh and Lewicki, James},
	month = apr,
	year = {2022},
}

@article{mardirossian_thirty_2017,
	title = {Thirty years of density functional theory in computational chemistry: an overview and extensive assessment of 200 density functionals},
	volume = {115},
	issn = {0026-8976},
	shorttitle = {Thirty years of density functional theory in computational chemistry},
	url = {https://doi.org/10.1080/00268976.2017.1333644},
	doi = {10.1080/00268976.2017.1333644},
	number = {19},
	urldate = {2025-11-07},
	journal = {Molecular Physics},
	author = {Mardirossian, Narbe and Head-Gordon, Martin},
	month = oct,
	year = {2017},
	pages = {2315--2372},
}

@article{eastwood_guidelines_2023,
	title = {Guidelines for designing peptoid structures: {Insights} from the {Peptoid} {Data} {Bank}},
	volume = {115},
	copyright = {© 2023 Wiley Periodicals LLC.},
	issn = {2475-8817},
	shorttitle = {Guidelines for designing peptoid structures},
	doi = {10.1002/pep2.24307},
	language = {en},
	number = {3},
	urldate = {2025-11-06},
	journal = {Peptide Science},
	author = {Eastwood, James R. B. and Weisberg, Ethan I. and Katz, Dana and Zuckermann, Ronald N. and Kirshenbaum, Kent},
	year = {2023},
	pages = {e24307}
}

@misc{simm_simpoly_2025,
	title = {{SimPoly}: {Simulation} of {Polymers} with {Machine} {Learning} {Force} {Fields} {Derived} from {First} {Principles}},
	shorttitle = {{SimPoly}},
	doi = {10.48550/arXiv.2510.13696},
	urldate = {2025-11-04},
	publisher = {arXiv},
	author = {Simm, Gregor N. C. and Hélie, Jean and Schulz, Hannes and Chen, Yicheng and Simeon, Guillem and Kuzina, Anna and Martinez-Baez, Ernesto and Gasparotto, Piero and Tocci, Gabriele and Chen, Chi and Li, Yatao and Cheng, Lixue and Wang, Zun and Nguyen, Bichlien H. and Smith, Jake A. and Sun, Lixin},
	month = oct,
	year = {2025},
	note = {arXiv:2510.13696 [physics]}
}

@article{wang_gaff2_2004,
	title = {Development and testing of a general amber force field},
	volume = {25},
	copyright = {Copyright © 2004 Wiley Periodicals, Inc.},
	issn = {1096-987X},
	url = {https://onlinelibrary.wiley.com/doi/abs/10.1002/jcc.20035},
	doi = {10.1002/jcc.20035},
	language = {en},
	number = {9},
	urldate = {2025-10-31},
	journal = {Journal of Computational Chemistry},
	author = {Wang, Junmei and Wolf, Romain M. and Caldwell, James W. and Kollman, Peter A. and Case, David A.},
	year = {2004},
	pages = {1157--1174},
}

@article{sansom_2019,
   author = {Marrink, Siewert J. and Corradi, Valentina and Souza, Paulo C. T. and Ingólfsson, Helgi I. and Tieleman, D. Peter and Sansom, Mark S. P.},
   title = {Computational Modeling of Realistic Cell Membranes},
   journal = {Chemical Reviews},
   volume = {119},
   number = {9},
   pages = {6184-6226},
   DOI = {10.1021/acs.chemrev.8b00460},
   year = {2019},
   type = {Journal Article}
}

@article{vattulainen_2019,
   author = {Enkavi, G. and Javanainen, M. and Kulig, W. and Rog, T. and Vattulainen, I.},
   title = {Multiscale Simulations of Biological Membranes: The Challenge To Understand Biological Phenomena in a Living Substance},
   journal = {Chem Rev},
   volume = {119},
   number = {9},
   pages = {5607-5774},
   ISSN = {1520-6890 (Electronic) 0009-2665 (Linking)},
   DOI = {10.1021/acs.chemrev.8b00538},
   url = {https://www.ncbi.nlm.nih.gov/pubmed/30859819},
   year = {2019},
   type = {Journal Article}
}

@InProceedings{oliver_beckstein-proc-scipy-2016,
  author    = { {R}ichard {J}. {G}owers and {M}ax {L}inke and {J}onathan {B}arnoud and {T}yler {J}. {E}. {R}eddy and {M}anuel {N}. {M}elo and {S}ean {L}. {S}eyler and {J}an {D}omański and {D}avid {L}. {D}otson and {S}ébastien {B}uchoux and {I}an {M}. {K}enney and {O}liver {B}eckstein },
  title     = { {M}{D}{A}nalysis: {A} {P}ython {P}ackage for the {R}apid {A}nalysis of {M}olecular {D}ynamics {S}imulations },
  booktitle = { {P}roceedings of the 15th {P}ython in {S}cience {C}onference },
  pages     = { 98 - 105 },
  year      = { 2016 },
  editor    = { {S}ebastian {B}enthall and {S}cott {R}ostrup },
  doi       = { 10.25080/Majora-629e541a-00e }
}

@article{samuli_2024,
   author = {Kiirikki, Anne M. and Antila, Hanne S. and Bort, Lara S. and Buslaev, Pavel and Favela-Rosales, Fernando and Ferreira, Tiago Mendes and Fuchs, Patrick F. J. and Garcia-Fandino, Rebeca and Gushchin, Ivan and Kav, Batuhan and Kučerka, Norbert and Kula, Patrik and Kurki, Milla and Kuzmin, Alexander and Lalitha, Anusha and Lolicato, Fabio and Madsen, Jesper J. and Miettinen, Markus S. and Mingham, Cedric and Monticelli, Luca and Nencini, Ricky and Nesterenko, Alexey M. and Piggot, Thomas J. and Piñeiro, Ángel and Reuter, Nathalie and Samantray, Suman and Suárez-Lestón, Fabián and Talandashti, Reza and Ollila, O. H. Samuli},
   title = {Overlay databank unlocks data-driven analyses of biomolecules for all},
   journal = {Nature Communications},
   volume = {15},
   number = {1},
   pages = {1136},
   ISSN = {2041-1723},
   DOI = {10.1038/s41467-024-45189-z},
   url = {https://doi.org/10.1038/s41467-024-45189-z},
   year = {2024},
   type = {Journal Article}
}

@ARTICLE{thompson2022lammps,
  title={LAMMPS-a flexible simulation tool for particle-based materials modeling at the atomic, meso, and continuum scales},
  author={Thompson, Aidan P and Aktulga, H Metin and Berger, Richard and Bolintineanu, Dan S and Brown, W Michael and Crozier, Paul S and int Veld, Pieter J and Kohlmeyer, Axel and Moore, Stan G and Nguyen, Trung Dac and others},
  journal={Computer Physics Communications},
  volume={271},
  pages={108171},
  year={2022},
  publisher={Elsevier}
}

@article{abbott2013polymatic,
  title={Polymatic: a generalized simulated polymerization algorithm for amorphous polymers},
  author={Abbott, Lauren J and Hart, Kyle E and Colina, Coray M},
  journal={Theoretical Chemistry Accounts},
  volume={132},
  number={3},
  pages={1334},
  year={2013},
  publisher={Springer}
}

@article{martinez2009packmol,
  title={PACKMOL: A package for building initial configurations for molecular dynamics simulations},
  author={Mart{\'\i}nez, Leandro and Andrade, Ricardo and Birgin, Ernesto G and Mart{\'\i}nez, Jos{\'e} Mario},
  journal={Journal of computational chemistry},
  volume={30},
  number={13},
  pages={2157--2164},
  year={2009},
  publisher={Wiley Online Library}
}

@article{maeda2016afir,
  title={Artificial Force Induced Reaction (AFIR) Method for Exploring Quantum Chemical Potential Energy Surfaces},
  author={Maeda, Satoshi and Harabuchi, Yu and Takagi, Makito and Taketsugu, Tetsuya and Morokuma, Keiji},
  journal={The Chemical Record},
  volume={16},
  number={5},
  pages={2232--2248},
  year={2016},
  publisher={Wiley Online Library}
}

@article{taylor2023architector,
  title={Architector for high-throughput cross-periodic table 3D complex building},
  author={Taylor, Michael G and Burrill, Daniel J and Janssen, Jan and Batista, Enrique R and Perez, Danny and Yang, Ping},
  journal={Nature Communications},
  volume={14},
  number={1},
  pages={2786},
  year={2023},
  publisher={Nature Publishing Group}
}

@misc{wood2025uma,
      title={UMA: A Family of Universal Models for Atoms}, 
      author={Brandon M. Wood and Misko Dzamba and Xiang Fu and Meng Gao and Muhammed Shuaibi and Luis Barroso-Luque and Kareem Abdelmaqsoud and Vahe Gharakhanyan and John R. Kitchin and Daniel S. Levine and Kyle Michel and Anuroop Sriram and Taco Cohen and Abhishek Das and Ammar Rizvi and Sushree Jagriti Sahoo and Zachary W. Ulissi and C. Lawrence Zitnick},
      year={2025},
      eprint={2506.23971},
      archivePrefix={arXiv},
      primaryClass={cs.LG},
      url={https://arxiv.org/abs/2506.23971}, 
}

@article{neese_software_2025,
	title = {Software {Update}: {The} {ORCA} {Program} {System}—{Version} 6.0},
	volume = {15},
	copyright = {© 2025 The Author(s). WIREs Computational Molecular Science published by Wiley Periodicals LLC.},
	issn = {1759-0884},
	shorttitle = {Software {Update}},
	doi = {10.1002/wcms.70019},
	language = {en},
	number = {2},
	urldate = {2025-05-11},
	journal = {WIREs Computational Molecular Science},
	author = {Neese, Frank},
	year = {2025},
	pages = {e70019},
}

@misc{yuan2025foundationmodelsatomisticsimulation,
      title={Foundation Models for Atomistic Simulation of Chemistry and Materials}, 
      author={Eric C. -Y. Yuan and Yunsheng Liu and Junmin Chen and Peichen Zhong and Sanjeev Raja and Tobias Kreiman and Santiago Vargas and Wenbin Xu and Martin Head-Gordon and Chao Yang and Samuel M. Blau and Bingqing Cheng and Aditi Krishnapriyan and Teresa Head-Gordon},
      year={2025},
      eprint={2503.10538},
      archivePrefix={arXiv},
      primaryClass={physics.chem-ph},
}

@article{acsmacrolett.5c00320,
author = {Adams, Cameron P. and Henein, Carolyn and Meng, Xiangxi and Yuan, Chenyun and Read de Alaniz, Javier and Ober, Christopher K. and Segalman, Rachel A.},
title = {Polymer Sequence Alters Sensitivity and Resolution in Chemically Amplified Polypeptoid Photoresists},
journal = {ACS Macro Letters},
volume = {14},
number = {8},
pages = {1055-1059},
year = {2025},
doi = {10.1021/acsmacrolett.5c00320},
}

\clearpage
\newpage

\let\addcontentsline\oldaddcontentsline
\appendix
\renewcommand{\contentsname}{Appendix Table of Contents}
\section*{Appendix}
\tableofcontents
\clearpage
\section{Calculation Details}\label{sec:app:dft_calcs}

All calculations in \op{} were carried out with the ORCA 6.0.0 DFT package~\cite{neese_software_2025}. ORCA supports various integral acceleration techniques, including RI-J and COSX, which dramatically improve the computational cost of these calculations at a very small cost in error, and these were used here. Experimentation with integral threshold settings indicated that the best trade-off between robust convergence and computational cost was to set the integral threshold (\texttt{thresh} in ORCA) to 1e-12 and the primitive batch threshold (\texttt{tcut} in ORCA) to 1e-13; these values were subsequently adopted by the ORCA package to be defaults in future versions of ORCA. ORCA's \texttt{tight} convergence settings were employed. Extensive benchmarking with various combinations of grid settings (both the exchange-correlation grid and the COSX grid) indicated that typical grids led to small numerical inconsistencies between energy and forces. In other words, there were discrepancies between the derivative of the energy with respect to coordinates and the computed forces due to grid incompleteness, and these errors were significant on the scale of errors with state-of-the-art MLIPs. In order to achieve sufficiently tight consistency, ORCA's \texttt{DEFGRID3} offered the best trade-off of convergence and cost. This corresponds to a pruned grid with 590 angular points for exchange-correlation and 302 for the final COSX grid.

\subsection{Calculation Quality Filters}\label{sec:app:quality}
We apply the same quality checks as were applied to the \op{} dataset. We enforce several quality checks on the resulting DFT calculations before considering them in the final dataset:

\begin{itemize}
    \item Referenced energies smaller in magnitude than $\pm$150 eV or with a referenced energy/atom of < 10 eV/atom. This removes highly unreasonable configurations; 
    \item Max per atom force not exceeding 50 eV/\angs. This also removes highly unreasonable configurations;
    \item $S^2 < 0.5$ for open-shell metal-containing systems, $S^2$ < 1.1 otherwise. We enforce tighter constraints on metal-systems to avoid incorrect SCF solutions but allow full spin-unpairing which may occur in organic reactivity;
    \item Enforce $\alpha$, $\beta$, and total electron consistency with the integrated densities. This indicates insufficient grid density;
    \item ORCA errors where the final COSX exchange deviates considerably. This indicate convergence errors;
    \item Non-negative HOMO-LUMO gaps.
\end{itemize}

%Furthermore, using a trained model on an early snapshot of OMol25, we perform an error analysis similar to the work of MACE-OFF on SPICE~\cite{kovacs2023mace}. Energy errors were evaluated for a held-out test set of \op{}, followed by manual inspection on some of the worst offenders. This allowed us to identify and filter systematically problematic inputs, such as isolated metal centers far away from the rest of the structure resulting in convergence to excited SCF minima.

\subsection{Computed Properties}\label{sec:app:calc}

The properties computed for each point in \op{} are the following:
\begin{enumerate}
    \item Total energy (in eV)
    \item Forces (in eV/\angs)
    \item charge
    \item spin
    \item Number of atoms
    \item Number of electrons
    \item Number of ECP electrons
    \item Number of basis functions
    \item Unrestricted vs. Restricted
    \item Number of SCF steps
    \item Energy computed by VV10
    \item S$^2$ expectation value
    \item Deviation of S$^2$ from ideal
    \item Integrated density (should be very close to the total number of electrons)
    \item HOMO energy (in eV), $\alpha$ and $\beta$ for unrestricted
    \item HOMO-LUMO gap (in eV), $\alpha$ and $\beta$ for unrestricted
    \item Maximum force magnitude for a given atom in a given direction (fmax)
    \item Mulliken charges (and spins if unrestricted)
    \item Loewdin charges (and spins if unrestricted)
    \item NBO charges (and spins if unrestricted) if the total number of atom <= 70
    \item Any ORCA warnings that are generated
    \item ORCA .gbw files and densities will be made available in the near future
\end{enumerate}

\section{Polymer Compositions}
\label{sec:app:polymer_compositions}

\subsection{General Sanitization}
\label{sec:app:smiles_sanitization}
All polymer repeat units are represented by a polymer SMILES string, such as '*CC*' for polyethylene. SMILES representations of polymer repeat units are sanitized by i) verifying that the SMILES is valid with the RDKit package,\cite{rdkit} and ii) enforcing that all polymer SMILES strings contain exactly two '*' characters, corresponding to the two points of connection between repeat units. To ensure that a representative number of repeat units are present in each simulation cell, we filter out any polymer repeat units that contain more than 50 total atoms. This choice ensures that, for example, the alternating copolymer with 150 atoms/chain have at least an A-B-A copolymer chain architecture.

\subsection{Traditional Polymers} \label{secsec:app:trad_polymers}
\begin{figure}[t]
    \centering
    \includegraphics[width=0.50\linewidth]{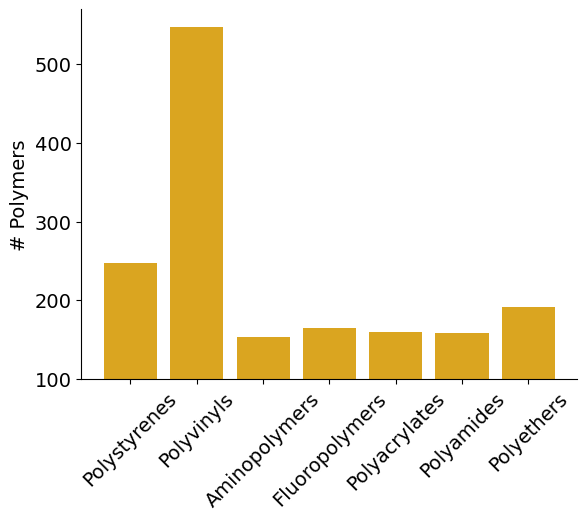}
    \caption{Overview of the polymer families present in the Traditional Polymer compositions. Polymer families are determined by defining a SMARTS substructure pattern for each polymer family and checking if each Traditional Polymer composition matches this pattern. }
    \label{fig:trad-polymer-compositions}
\end{figure}

The majority of the traditional polymer SMILES strings are sourced from the RadonPy benchmark dataset of 1,077 unique homopolymers.\cite{hayashi_radonpy_2022} According to the RadonPy paper, these 1,077 polymers were selected from the PoLyInfo database,\cite{otsuka_polyinfo_2011} which contains 15,335 homopolymers, based on having the largest number of recorded experimental property measurements. To provide a further diversity of polymer compositions, we also hand-collected 85 unique homopolymers from the Bicerano Polymer Handbook\cite{bicerano_prediction_2002}. Filtering and sanitizing these SMILES strings yields our final dataset of 840 traditional polymer compositions. In Figure \ref{fig:trad-polymer-compositions}, we highlight the general polymer families present in the Traditional Polymer compositions.  

\subsection{Fluoropolymers}
Due to the lack of open-source databases of experimentally synthesized fluoropolymer SMILES strings, we instead collect example fluoropolymer compositions from the OpenMacromolecularGenome (OMG).\cite{kim_OMG_2023} OMG consists of approximately 12 million unique polymer compositions that have been generated by applying 17 known polymer template reactions (such as step growth or chain growth addition) to a curated list of 77,281 commercially available small molecule reactants. Although all of the polymer compositions in OMG are not guaranteed to be synthesizable, this polymer generation process helps to ensure that compositions in OMG can be attributed to a proposed synthetic pathway.

\begin{figure}[t]
    \centering
    \includegraphics[width=1.0\linewidth]{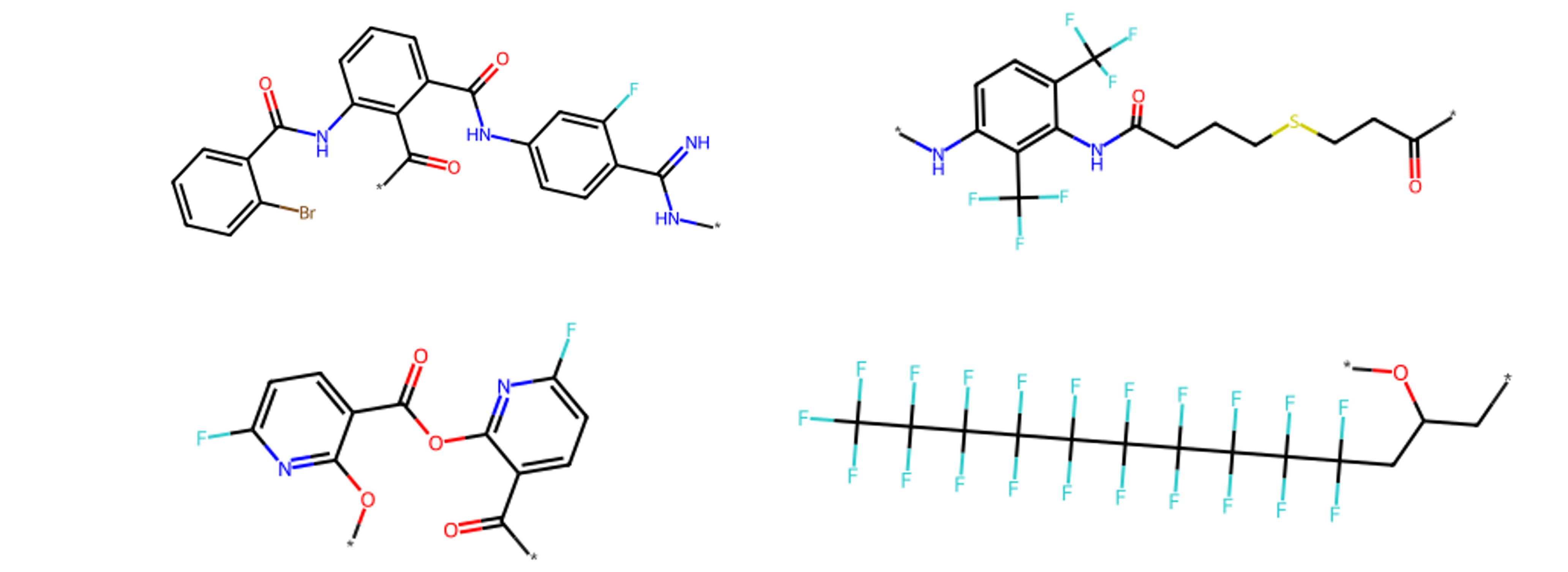}
    \caption{Illustrative examples of the fluoropolymer compositions present in \op{}, depicting polymers with a wide range of fluorine content.}
    \label{fig:fluoropolymer-compositions}
\end{figure}

From this dataset of 12 million polymer compositions, we first filter the compositions to check for SMILES validity (see \ref{sec:app:smiles_sanitization}) and identify any polymer composition with at least one F atom, resulting in approximately 2.5 million fluoropolymer compositions. Then, to ensure that we sample as diverse a range of fluoropolymers as possible, we separate each of the fluoropolymers into their 17 associated polymer reaction templates. Within each polymer reaction template, we sort the polymer compositions by their fluorine content (as a fraction of total number of atoms) and then uniformly sample from the lower, middle, and upper thirds of this distribution. Since the fraction of fluorine in a polymer strongly influences its electronic and physical properties, this sampling strategy serves to ensure that \op{} contains polymers with a diverse range of fluorine content. Similarly, uniformly sampling across all 17 polymer reaction templates ensures that we capture polymer compositions synthesized by various synthetic pathways. Altogether, we obtain 521 unique fluoropolymer compositions. We depict several illustrative fluoropolymer structures in Figure \ref{fig:fluoropolymer-compositions}

\subsection{Optical Polymers}
We obtain optical polymer compositions from the chemical space of conjugated polymers provided by Aldeghi et al.\cite{aldeghi_graph_2022}, which was originally experimentally explored by Bai et al.\citep{bai_accelerated_2019} Notably, this conjugated polymer library consists of prescribed copolymers that are comprised of 9 diboronic acid/acid ester (A monomers) and 706 dibromo (B) monomers, resulting in a total library of 6354 candidate copolymers. In this work, we respect these original A/B monomer assignments such that all copolymers are formed by the combination of one A monomer with one B monomer.

\subsection{Polymer Electrolytes}
All polymer electrolyte compositions are obtained from the PolyGen dataset of 6,024 unique polymers\cite{xie_accelerating_2022,lei2023self,yang2023novo}. Notably, these polymer compositions are based off of hypothetical homopolymers that could be synthesized from a known condensation polymerization route. After sanitizing these compositions, we then represent all compositions with RDKit descriptors, (implemented through the \textit{RDKitDescriptors} class of the DeepChem package)\cite{Ramsundar-et-al-2019}. In order to sample the most diverse monomers from this dataset, we then perform k-means clustering on these featurized monomers (with k=10) and randomly sample 30 monomer compositions to give 300 unique polymer electrolyte compositions.

\subsection{Ion-Inserted Polymers} \label{secsec:app:ion_inserted_polymers}
Ion-inserted polymer systems were constructed as follows. Equilibrated polymer configurations in PDB format, obtained from prior classical molecular dynamics simulations, were used as the starting structures. For each polymer, a set of twenty ions was randomly selected from a curated library of forty common monoatomic and polyatomic ions (see listing in Table \ref{tab:ions}).

The selected ions were inserted into the equilibrated polymer matrices using the \textsc{Gromacs} molecular simulation package (version~2023.4). The \texttt{insert-molecules} utility was employed to place each ion template into the simulation box through stochastic trial placements, automatically rejecting overlapping configurations. This ensured physically realistic ion--polymer packing without manual adjustments. After insertion, all generated PDB files were standardized by correcting element symbols and assigning formal residue charges based on canonical oxidation states, with special handling for polyatomic species to preserve their stoichiometric and charge balance. For multi-atom ions, internal bonding information (\texttt{CONECT} records) was reconstructed from reference templates and appended to each combined polymer--ion structure while retaining the original polymer connectivity. The resulting outputs consisted of validated, charge-balanced PDB systems containing polymers with randomly distributed ions, complete connectivity information, and consistent metadata. These systems were then subjected to 2000 steps of MLIP-MD with UMA-s-1 OMol task, after which ion-containing subsystems were extracted for DFT simulations.

\subsection{Lipids}
\begin{figure}[H]
    \centering
    \includegraphics[width=0.9\linewidth]{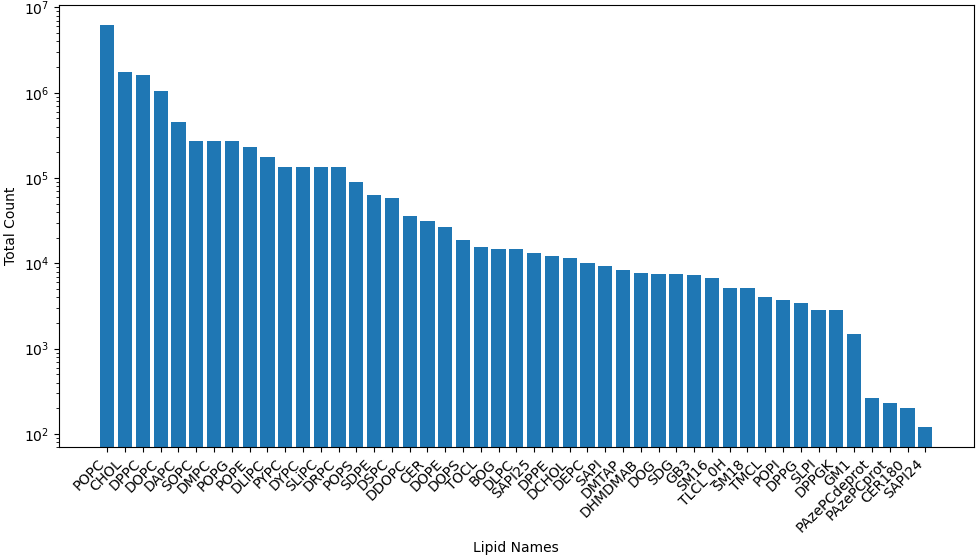}
    \caption{Total lipid counts across all simulations included in OPoly26.}
    \label{fig:lipid_types}
\end{figure}

To span a diverse set of common lipids in different bilayer environments, we extracted lipid simulation snapshots from the NMRlipid database \cite{samuli_2024}. We sampled simulations at 10 ns interval using all simulations listed in the NMRlipid database fitting with the following criteria: simulated with one of the following force fields (CHARMM, Slipids, lipid14, lipid17, GROMOS, OPLS, Berger), longer than 20~ns, parsable with MDAnalysis \cite{oliver_beckstein-proc-scipy-2016}, and contained atom charge information. The final set contained 47 different lipid types sampled from 674 simulations. Figure \ref{fig:lipid_types} shows the total number of lipids of each type across all the simulations.

\subsection{Polymer Architectures}
As inputs, RadonPy takes in a SMILES string representation of the repeat unit sequence, the number of atoms in each polymer chain, the polymer type (including homopolymer, alt. copolymer, or random copolymer) and the number of polymer chains in the simulation cell. RadonPy then utilizes a self-avoiding random walk procedure to produce initial bulk amorphous cells. All chains are treated as atactic, and chain architectures are limited to linear homopolymers and copolymers, including alternating and random copolymers. 
\section{MD Details}
\label{sec:app:md_details}

To generate large quantities of polymer structures, we carried out a series of classical all-atom MD simulations using the LAMMPS MD suite\cite{thompson2022lammps} and the RadonPy tools made available by Hayashi et al.\cite{hayashi_radonpy_2022} To facilitate the automation of MD calculations, RadonPy generates and runs LAMMPS input decks. 

Classical MD calculations use fully periodic boundary conditions (PBCs). Standard Velocity Verlet time integration, with a timestep of 1 fs is used for MD production runs. This timestep is paired with the application of the SHAKE to all bonds with hydrogens (effectively fixing the lengths of these bonds). Where applicable, a Nos\'e-Hoover thermostat and barostat are applied with time constants of 100 and 1000 fs respectively.

System preparation procedures (detailed in the SI) leverage a “21-step” compression-decompression procedure proposed by Abbott et al.,\cite{abbott2013polymatic} that uses alternating NPT (isothermal-isobaric) and high-temperature NVT (isothermal-isochoric) annealing phases to accelerate polymer dynamics (assuming time-temperature superposition). 
We opt to anneal the system at 600 K. This approach accelerates the equilibration of polymer melt configurations and defines a physically meaningful correlation time when sampling pseudo-independent simulation frames.
%We monitor chain relaxation dynamics during equilibration and production runs, adjusting the annealing temperature (either 600 K or 1000 K) as needed. The benefit of this is two-fold, from an equilibration perspective and from the perspective of producing pseudo-“independent” or decorrelated frames over time.
%NTL note: ran out of time to do a second set of calculations at 1000 K
As a final caveat, note that in a few cases 600 K may be insufficient to produce independent frames (e.g., polyimides); chain relaxation dynamics can be quite slow, and chain length dependent ($\tau \propto N^2$ up to $\tau \propto N^{3.4}$).

All classical MD uses a modified version of the General Amber FF (GAFF2),\cite{wang_gaff2_2004,trag_improved_gaff2_2019} which is a generic atomistic potential, expressed in a convenient class 1 style. Specifically, GAFF2 includes harmonic bonds and angles, Fourier-style proper dihedrals, and cosine-style improper dihedrals. Van der Waals interactions are expressed in the standard 12-6 Lennard-Jones form with standard Lorentz-Bertholot mixing rules and incorporate switching functions to smoothly approach zero at 12\AA. Coulomb interactions were evaluated in real space up to an 12\AA cutoff and interactions beyond the cutoff were evaluated with the PPPM k-space solver using a relative error tolerance of 1e-6. Atomic electric charges were assigned using RadonPy\cite{hayashi_radonpy_2022} and were computed using the RESP method\cite{bayly_well-behaved_1993} and electrostatic potentials derived from DFT calculations performed at the ωB97M-D3BJ level on isolated, hydrogen-capped monomers.\cite{mardirossian_b97m-v_2016} Standard Amber intramolecular pairwise exclusion rules were applied in which 1-4 intramolecular interactions were scaled by 0.5 for LJ terms and 0.833 for Coulomb terms. Further FF parameterization details (including charge assignment, cutoffs, heteratom mixing, and intramolecular exclusion rules) are covered in \ref{SI:md-ffield}, and the original RadonPy\cite{hayashi_radonpy_2022} manuscript.

\subsection{MD Force Field}\label{SI:md-ffield}

All calculations use a modified version of the General Amber FF, GAFF2, which is a generic atomistic potential.\cite{wang_gaff2_2004, trag_improved_gaff2_2019}.
As highlighted above, intramolecular potential energies are expressed in a convenient class 1 style; all bonds and angles are treated harmonically. The Fourier-style proper dihedrals (with angle $\phi$) can be expressed via,

\begin{equation}
E = \sum_{i=1}^{m}{K_i\big[1+\cos\big(n\phi-d_i\big)\big]}
\end{equation}

\noindent while improper dihedrals are expressed via the cosine form, i.e.,

\begin{equation}
E = K\big[1+d\cos(n\phi)\big].
\end{equation}

%\noindent\todo{This appears to be at odds with Eq.~2 in Ref.~}\citenum{hayashi_radonpy_2022}\todo{, but this is what the RadonPy package uses via the generated LAMMPS input decks.}

VDW interactions consistently follow a standard 12-6 Lennard-Jones form. However, there is a short period of initial relaxation dynamics (during which coulombic interactions are switched off) that applies a simple radial cutoff of 3 angstroms to the 12-6 LJ potential. All subsequent runs apply a CHARMM-style switching function to the LJ potential that ramps the energy smoothly to zero between 8 and 12 angstroms (rather than having a discrete radial cutoff distance). The proposed benefit of this switching function is to reduce artifacts associated with the discontinuous change in the intermolecular potential at a discrete radial cutoff. Finally, we note that heteroatom mixing is arithmetic for LJ diameters, and geometric for the interaction strength parameter.

Intramolecular exclusion rules from Amber are adopted, meaning the LJ and coulombic interactions exclude atoms on the same molecule separated by 1 or 2 bonds (known as 1-2 and 1-3 interactions). Following the amber FF defaults, 1-4 interactions are scaled down by a factor of 0.5 for LJ, and 5/6 for coulombic interactions.

Coulombic interactions are handled differently; real-space interactions use a fixed cutoff of 12 \AA (i.e., no switching functions are applied), whereas long-range electrostatics are handled via the P$\rm^3$M algorithm, which involves mapping atomic charges onto a fine mesh or grid. P$\rm^3$M settings are configured to target a per-atom force tolerance (quantified by the RMS error divided by a reference force, representing the pairwise interaction of two monovalent ions) of 10$^{-6}$ (dimensionless). All other long-range solver settings are the LAMMPS suite defaults. Atomic partial charges are set is described in more detail below, as well as in the RadonPy paper.\cite{hayashi_radonpy_2022} 
To ensure pair interactions aren’t missed, neighbor lists are updated on each step in which an atom moves more than half the skin distance (set to 2 \AA).

\subsection{System Preparation}

System preparation follows the details provided in the RadonPy paper; in short this is a multistage system preparation procedure, where:

\newcounter{tempcounter}
\begin{enumerate}
\item RDKit's conformer generator (ETKDG version 2) generates the initial molecular topology of a family of H-terminated repeat units from a SMILES string, which are
\item geometry optimized / energy minimized via a Molecular Mechanics calculation with the General Amber FF and generic charges.
\item The 4 most energetically stable conformers are then selected and DFT-optimized, and
\item fixed atomic charges are calculated using the RESP charge scheme, ensuring charge neutrality.
\item Individual chains are then polymerized from these repeat units via a self-avoiding random walk algorithm.
\setcounter{tempcounter}{\value{enumi}}
\end{enumerate}

\noindent Options are provided to generate homopolymers, random, alternating, or block copolymers as random walks. Options are exposed in the RadonPy suite to specify fractions of co-monomers. Tacticity (atactic, isotactic, syndiotactic) is available as a secondary option. All polymer chains considered in our calculations are treated as atactic. All polymer chains are capped with a hydrogen termination group.

\subsubsection{Preparation of Bulk (Amorphous) Polymer Cells}

We opt to prepare two sets of cells --- primarily, we focus on preparing boxes with 10 (approx.) 500-atom chains. In practice, this means that the degree of polymerization of our  polymers (and in many cases, really oligomers) is variable, to keep the number of atoms per chain fixed. We note that these cells are approximately half the size of the 10000 atom cells prepared in the RadonPy manuscript. Having said that, we do prepare very small supplemental oligomer cells with just 3 100 atom chains; these cells are prepared only because it is impractical to run high-throughput DFTB MD calculations (described in a later section) on larger cells.

Subsequently, there is an initial packing procedure (described fully in Ref.~\citenum{hayashi_radonpy_2022}) involving:

\begin{enumerate}
 \setcounter{enumi}{\value{tempcounter}}
\item Creating an initial low density (0.05 g/cm$\rm ^3$) structure by translating and rotating N chains, placing each chain into the box such that there are no overlaps.
\item Geometry optimization is then used to relax the initial configuration, followed by
\item 20 ps of 300 K NVT dynamics, with a reduced timestep (0.1 fs), and 
\item 1 ns of dynamics with a NH thermostat, while temperature is ramped from 300 K to 700 K.
\item Finally, 1 ns of 700 K dynamics is carried out with a NH thermostat, while the cell is deformed until a density of 0.8 g/cm$\rm ^3$ is achieved.
\end{enumerate}

Steps (7-9) neglect coulombic interactions, but retain LJ interactions, with a fixed radial cutoff of 3 angstroms (in contrast, to the production runs detailed above). For more granular detail, refer to the RadonPy manuscript,\cite{hayashi_radonpy_2022} where each of these steps is described in full. Subsequently, we use a modified version of the equilibration schedule proposed by Larsen and co-workers. This procedure is sometimes referred to as a “21 step” procedure, which makes the procedure sound unnecessarily complicated.
In short, it boils down to a multi-stage annealing procedure, consisting of a compression and decompression phase, with 3 discrete steps up (0.1, 3, 5), and 4 steps down (2.5, 0.5, 0.05 GPa, 1 atm) in pressure. 

After each incremental step up (or down) in pressure: 
\begin{enumerate}
\item 300 K NPT dynamics are run with a barostat enforcing the higher (or lower) pressure.
\item Next, there is an annealing step, with high-temperature NVT dynamics (at 600 K or 1000 K, depending on the speed of the chain’s dynamics).
\item The system is re-thermalized via 300 K NVT dynamics; a NH thermostat cools the system down and enforces the target temperature, prior to continuing with the compression-decompression cycle.
\end{enumerate}

The principle of step 2 is to accelerate chain relaxation, based on a time-temperature superposition type assumption. On this basis, we choose to increase the length of the high-temperature NVT dynamics stages during the decompression phase; specifically, we use 50 ps of 600 K (or 1000 K) of NVT dynamics during the compression phase, whereas 500 ps runs are used during decompression. The final 300 K, 1 atm target state was also lengthened, using 1 ns of high-temperature annealing, a 10 ps re-thermalization run (NVT at 300 K), and then 800 ps of 300 K, 1 atm NPT dynamics. For a full equilibration schedule, see below.

\begin{table}[h]
\begin{center}
\caption{Equilibration schedule detailing each of the 21 steps in the compression-decompression cycle used to prepare bulk amorphous polymer cells.}
\begin{tabular}{|c|c|c|c|}
\hline 
Cycle \# & Ensemble & T, P & Time {[}ps{]}\tabularnewline
\hline 
\hline 
\multirow{3}{*}{1} & \multirow{2}{*}{NVT} & $T_{{\rm max}}=$ 600 K (or 1000 K) & 50\tabularnewline
\cline{3-4} \cline{4-4} 
 &  & 300 K & 50\tabularnewline
\cline{2-4} \cline{3-4} \cline{4-4} 
 & NPT & 300 K, .1 GPa & 50\tabularnewline
\hline 
\multirow{3}{*}{2} & \multirow{2}{*}{NVT} & 600 K (or 1000 K) & 50\tabularnewline
\cline{3-4} \cline{4-4} 
 &  & 300 K & 100\tabularnewline
\cline{2-4} \cline{3-4} \cline{4-4} 
 & NPT & 300 K, 3 GPa & 50\tabularnewline
\hline 
\multirow{3}{*}{3} & \multirow{2}{*}{NVT} & 600 K (or 1000 K) & 50\tabularnewline
\cline{3-4} \cline{4-4} 
 &  & 300 K & 100\tabularnewline
\cline{2-4} \cline{3-4} \cline{4-4} 
 & NPT & 300 K, $P_{{\rm max}}=$ 5 GPa & 50\tabularnewline
\hline 
\multirow{3}{*}{4} & \multirow{2}{*}{NVT} & 600 K (or 1000 K) & 500\tabularnewline
\cline{3-4} \cline{4-4} 
 &  & 300 K & 100\tabularnewline
\cline{2-4} \cline{3-4} \cline{4-4} 
 & NPT & 300 K, 2.5 GPa & 5\tabularnewline
\hline 
\multirow{3}{*}{5} & \multirow{2}{*}{NVT} & 600 K (or 1000 K) & 500\tabularnewline
\cline{3-4} \cline{4-4} 
 &  & 300 K & 10\tabularnewline
\cline{2-4} \cline{3-4} \cline{4-4} 
 & NPT & 300 K, 0.5 GPa & 5\tabularnewline
\hline 
\multirow{3}{*}{6} & \multirow{2}{*}{NVT} & 600 K (or 1000 K) & 500\tabularnewline
\cline{3-4} \cline{4-4} 
 &  & 300 K & 10\tabularnewline
\cline{2-4} \cline{3-4} \cline{4-4} 
 & NPT & 300 K, 0.05 GPa & 5\tabularnewline
\hline 
\multirow{3}{*}{7} & \multirow{2}{*}{NVT} & 600 K (or 1000 K) & 1000\tabularnewline
\cline{3-4} \cline{4-4} 
 &  & 300 K & 10\tabularnewline
\cline{2-4} \cline{3-4} \cline{4-4} 
 & NPT & 300 K, 0 GPa (1 atm) & 800\tabularnewline
\hline 
\end{tabular}
\end{center}
\end{table}

\subsubsection{Preparation of Solvated (Infinite Dilution) Polymer Cells} \label{sssec:app:solvated_cells}

Due to practical time constraints, we opt to prepare a series of infinite dilution solvated cells; specifically consisting of a single 500 atom chain, and 4500 solvent atoms (although the established pipeline can prepare multi-chain solvated cells with some limitations). A full list of solvents and polymers is provided in the appendix.
As with our bulk cells, the GAFF2 FF is used to represent both polymer and solvent at atomistic resolution. The sole exception to this is water, which we use with a 3-site flexible water model (TIP3P/FW) that was modified for use with long-range (Ewald style) electrostatics.\cite{price_tip3p_modified_2004} Exact parameters for the water model are provided in \ref{sec:app:solvents}. Initial preparation stages are identical to those outlined previously. As with the bulk cells, RESP charges are derived from DFT-level calculations of the polymer’s constituent repeat units and solvent molecules respectively.
After using RadonPy to polymerize a single chain (homopolymer, alternating copolymers, or random copolymer) via a self-avoiding random walk, we stamp one of a handful of chosen solvent molecules into a periodic box using Packmol.\cite{martinez2009packmol} Packmol tries to neatly pack the solvent molecules around the lone polymer chain (representing a chain at infinite dilution) such that it achieves the desired target density of 1 g/cm$\rm ^3$ we selected. Since Packmol is entirely ignorant of any FF parameters or information, we use a

\begin{enumerate}
\item preliminary energy minimization of the Packmol cell to clean up the initial configurations. To ensure the configuration is well-behaved, we subsequently
\item do an initial NVT relaxation at a lower temperature (100 K). A reduced timestep (0.1 and then 0.5 fs) is used with intermittent velocity reselects from a  Maxwell-Boltzmann distribution.
\item Nose-Hoover style temperature and pressure control is used to gradually ramp the temperature up from 100 K to the target of 300 K at 1 atm.
\item NVT relaxation at 300 K with a few velocity reselects, followed by
\item NPT relaxation at 300 K and 1 atm.
\item 200 ps of high-temperature NVT dynamics are used (at 600 K).
\item Ramp down temperature to 300 K, followed by NVT thermalization at 300 K.
\item A final period of NPT equilibration at 300 K and 1 atm (200 ps)
\end{enumerate}

\subsection{Solvents}\label{sec:app:solvents}

\begin{table}[h!]
\centering
\caption{Solvents used in MD polymer simulations. The density values represent the inputs densities used to construct the initial MD simulation cells. Water was simulated with the TIP3P(Ewald) FF\cite{price_tip3p_modified_2004}. All other solvents were simulated with the GAFF2 FF\cite{wang_gaff2_2004} with modified fluorocarbon parameters\cite{trag_improved_gaff2_2019}. Out-of-distribution (OOD) solvents were not included in either the train set of \op{} nor OMol25.}
\begin{tabular}{l l l l l}
\hline
\textbf{Solvent Name} & \textbf{SMILES} & \textbf{Density (g/cm$^3$)} & FF Parameters & Train/OOD\\
\hline
Water & O & 1.00 & TIP3P\cite{price_tip3p_modified_2004} & Train\\
Acetone & CC(=O)C & 0.791 & GAFF2\cite{wang_gaff2_2004,trag_improved_gaff2_2019} & Train\\
Acetonitrile & CC\#N & 0.786 & GAFF2\cite{wang_gaff2_2004,trag_improved_gaff2_2019} & Train\\
Benzene & C1=CC=CC=C1 & 0.876 & GAFF2\cite{wang_gaff2_2004,trag_improved_gaff2_2019} & Train\\
Cyclohexane & C1CCCCC1 & 0.779 & GAFF2\cite{wang_gaff2_2004,trag_improved_gaff2_2019} & Train\\
Hexane & CCCCCC & 0.661 & GAFF2\cite{wang_gaff2_2004,trag_improved_gaff2_2019} & Train\\
Octane & CCCCCCCC & 0.703 & GAFF2\cite{wang_gaff2_2004,trag_improved_gaff2_2019} & Train\\
Propanol & CCCO & 0.803 & GAFF2\cite{wang_gaff2_2004,trag_improved_gaff2_2019} & Train\\
1,2,4- Trichlorobenzene & C1=CC(=C(C=C1Cl)Cl)Cl & 1.46 & GAFF2\cite{wang_gaff2_2004,trag_improved_gaff2_2019} & Train\\
Dimethyl Formamide & CN(C)C=O & 0.95 & GAFF2\cite{wang_gaff2_2004,trag_improved_gaff2_2019} & Train\\
Tetrahydrofuran & C1CCOC1 & 0.888 & GAFF2\cite{wang_gaff2_2004,trag_improved_gaff2_2019} & Train\\
Chloroform & C(Cl)(Cl)Cl & 1.49 & GAFF2\cite{wang_gaff2_2004,trag_improved_gaff2_2019} & Train\\
Phenol & C1=CC=C(C=C1)O & 1.07 & GAFF2\cite{wang_gaff2_2004,trag_improved_gaff2_2019} & Train\\
Dichloromethane & C(Cl)Cl & 1.33 & GAFF2\cite{wang_gaff2_2004,trag_improved_gaff2_2019} & Train\\
Toluene & CC1=CC=CC=C1 & 0.867 & GAFF2\cite{wang_gaff2_2004,trag_improved_gaff2_2019} & Train\\
Methanol & CO & 0.792 & GAFF2\cite{wang_gaff2_2004,trag_improved_gaff2_2019} & Train\\
n-Heptane & CCCCCCC & 0.684 & GAFF2\cite{wang_gaff2_2004,trag_improved_gaff2_2019} & Train\\
1,4-Dioxane & O1CCOCC1 & 1.033 & GAFF2\cite{wang_gaff2_2004,trag_improved_gaff2_2019} & OOD\\
Formic Acid & O=CO & 1.220 & GAFF2\cite{wang_gaff2_2004,trag_improved_gaff2_2019} & OOD\\
N-Methyl-2-Pyrrolidone & CN1CCCC1=O & 1.028 & GAFF2\cite{wang_gaff2_2004,trag_improved_gaff2_2019} & OOD\\
Decalin & C1CCC2CCCCC2C1 & 0.896 & GAFF2\cite{wang_gaff2_2004,trag_improved_gaff2_2019} & OOD\\
Tetrachloroethylene & ClC(Cl)=C(Cl)Cl & 1.622 & GAFF2\cite{wang_gaff2_2004,trag_improved_gaff2_2019} & OOD\\
\hline
\end{tabular}
\end{table}

\subsection{Ion insertion MLIP-MD}\label{sec:app:ion_inserted_mlip_md}

\begin{table}[h!]
\centering
\caption{Ions used in MLIP-MD simulations. Test ions were not included in the train set of \op{}, but may be present in OMol25.}
\label{tab:ions}
\begin{tabular}{l l l}
\hline
\textbf{Ion Name} & \textbf{Chemical Formula} & Train/Test\\
\hline
Aluminum Ion & $Al^{3+}$ & Train\\
Tetrafluoroborate & $BF_4^{-}$ & Train\\
Bromide& $Br^{-}$ & Train\\
Calcium Ion & $Ca^{2+}$ & Train\\
Triflate & $CF_3SO_3^{-}$ & Train\\
Acetate & $CH_3COO^-$ & Train\\
Perchlorate& $ClO_4^{-}$ & Train\\
Chloride & $Cl^{-}$ & Train\\
Cyanide & $CN^-$ & Train\\
Cobalt Ion & $Co^{2+}$ & Train\\
Carbonate & $CO_3^{2-}$ & Train\\
Cesium Ion & $Cs^{+}$ & Train\\
Copper Ion & $Cu^{2+}$ & Train\\
Iron Ion & $Fe^{2+}$ & Train\\
Fluoride & $F^{-}$ & Train\\
Bicarbonate & $HCO_3^{-}$ & Train\\
Iodide & $I^{-}$ & Train\\
Potassium Ion & $K^{+}$ & Train\\
Lanthanum Ion & $La^{3+}$ & Train\\
Lithium Ion & $Li^{+}$ & Train\\
Magnesium Ion & $Mg^{2+}$ & Train\\
Sodium Ion & $Na^{+}$ & Train\\
Ammonium & $NH_4^{+}$ & Train\\
Nickel Ion & $Ni^{2+}$ & Train\\
Nitrate & $NO_3^{-}$ & Train\\
Hexafluorophosphate & $PF_6^{-}$ & Train\\
Phosphate & $PO_4^{3-}$ & Train\\
Sulphate & $SO_4^{2-}$ & Train\\
Strontium Ion & $Sr^{2+}$ & Train\\
Zinc Ion & $Zn^{2+}$ & Train\\
Hydronium & $H_3O^+$ & Test\\
Hypochlorite & $ClO^-$ & Test\\
Sulfite & $SO_3^{2-}$ & Test\\
Thiocyanate & $SCN^-$ & Test\\
\hline
\end{tabular}
\end{table}

The  high-entropy copolymer and peptoid systems equilibrated with classical MD underwent ion-insertion and  were subsequently used as the initial configurations for machine-learning molecular dynamics (ML-MD) simulations. All simulations were performed using the Atomic Simulation Environment (ASE) framework interfaced with the FAIRChem machine-learned interatomic potential (MLIP) package.\cite{Larsen_2017} Atomic interactions were modeled using the UMA-S-1 neural network potential, which was selected for its balanced accuracy across organic and inorganic chemistries relevant to ion–polymer interactions.\cite{wood2025uma} Each polymer–ion configuration was converted into an ASE Atoms object and simulated under periodic boundary conditions applied in all three directions. The total formal charge of each system was automatically parsed from the PDB metadata and passed to the ML potential to maintain charge consistency. MD simulations were carried out in the canonical ensemble (NVT) at 300 K using a Berendsen thermostat with a relaxation time constant of 100 fs. The equations of motion were integrated with a 1 fs time step, and atomic velocities were initialized from a Maxwell–Boltzmann distribution at the target temperature. All production trajectories were run for 2 ps before extracting clusters for single point DFT calculations. 

\subsection{DFTB} \label{sec:app:dftb}
We enriched the configuration space sampled for the set of Traditional Homopolymers through semiempirical quantum MD applied to simulate both reactive and nonreactive conditions. To this end, we simulate polymers with the third-order Density Functional Tight Binding (DFTB) method,\cite{gaus_dftb3_2011} the DFTB+ code,\cite{hourahine_dftb_2020} and the 3ob-3-1 parameter set (available at http://www.dftb.org),\cite{gaus_parameterization_2014} which is a general parameter set for organic molecules. The electronic structure was evaluated at the gamma point without spin polarization and with Fermi-Dirac thermal smearing with the electronic temperature set equal to the ionic thermostat temperature.\cite{mermin_thermal_1965} DFTB-MD simulations were performed with NVT dynamics at 300 K and 600 K using 3D periodic equilibrium amorphous configurations (<300 atoms) prepared with classical MD through RadonPy. Simulations were run for at least 20,000 time steps, with a 0.2 fs time step. A small set of highly reactive simulations probing radiation-induced chemistry of silicones were performed following the protocols in Ref. \cite{kroonblawd_polymer_2022}.
\section{Reactivity}\label{sec:app:react_details}

Diversity in reactive configurations, with minimal bias toward specific chemical motifs, is ensured by selecting a single bond for dissociation randomly from a candidate pool curated to mitigate C–H bond overrepresentation and exclude nonphysical pathways involving aromatic atoms. By avoiding prespecified product structures, the AFIR search remains unbiased toward known reaction outcomes, allowing the discovery of reaction pathways over the course of data generation. We apply this protocol to homopolymer and copolymer systems, as well as bulk and solvated chains. To introduce additional electronic diversity, 10\% of chains are protonated and 10\% deprotonated near the reactive bond, and 10\% undergo single-electron oxidation and 10\% single-electron reduction (AFIR generation of these radical systems had a high failure rate, so a comparatively small sample of final data points in the dataset were derived from these reduced or oxidized reactions). These procedures generate rich training data with local environments where charge and unpaired electrons can modulate reactivity in these polymer system.  

To more faithfully capture the polymer chain environment, each AFIR trajectory is initialized from a unique chain conformation where an atom at each chain end is fixed in space. We impose these constraints because polymer chains, being macromolecular in nature, have limited access to other conformations due to the entropic constraint of being confined by sizable molecular fragments that make up an overall chain. After AFIR trajectories are generated, frames are subsampled by the same procedure put forward in OMol25\cite{levine2025openmolecules2025omol25} and trimmed to under 250 atoms as preparation for DFT calculations. With this procedure we obtain high chemical diversity across tens of thousands of single-bond dissociation events and incorporate approximations that capture the differences in chemistry between polymers and small molecules. 

\subsection{AFIR Preparation}\label{ssec:afir_prep}
Polymer chain structures from MD trajectories are encoded using a modified \texttt{Molecule} class in Architector\cite{taylor2023architector}, which contains additional attributes for repeat unit representations, solvent molecules, and chain end atoms. RDKit\cite{rdkit}, ASE, and OpenBabel facilitate the reconstruction of molecular connectivity and geometry from MD-generated structures, correcting periodic artifacts and broken bonds by comparing covalent radii with wrapped interatomic distances. When the indices of chain end atoms (i.e., those that share a bond with the hydrogen cap on the terminal repeat unit) are not explicitly specified in the MD structure file, chain termini are located with RDKit substructure matching. Metadata such as repeat unit(s) and solvent type are parsed from MD frame filenames and mapped to the curated SMILES library for that category (e.g., traditional, optical, fluoropolymer, electrolyte, solvent).

\subsubsection{Bond Selection}
Bonds are first scored and ranked based on local chemical environment metrics. A final bond is then selected at random from a filtered, high-scoring subset of bonds, allowing stochastic diversity in bond selection while baking in chemical intuition. 

Polymers are predominantly hydrocarbon macromolecules. 
To address this inherent skew in bond-type distribution in the interest of exploring a diverse set of bond dissociation reactions in polymer systems, we enforce that only 10\% of the subset of high-scoring bonds contain hydrogen. Additionally, we limit ring-containing bonds to be at most 10\% of the subset, in order to prevent an over-selection of such bonds when scoring is based on local atomic density. In no case, however, are aromatic bonds present in the subset. Dissociation reactions that involve aromatic bonds are likely to lead to convergence failures under standard DFT protocols, and are not frequently studied reactions due to their incredibly low probability of occurrence. Bond dissociation reactions that result in forming triplet oxygen are also similarly avoided.

For bulk versus solvated chain systems, we employ differently tailored bond scoring procedures: 

\textbf{Bulk reactivity:} In bulk systems, we are interested in reactivity at the point of highest density to stay faithful to the condensed-phase environment of bulk polymer. Candidate bonds are therefore required to be (a) at least 5.0 Å away from all chain ends (as defined in \ref{ssec:afir_prep}) and (b) within 5.0 Å of the polymer system's center of mass. If no such bond satisfies these criteria, which can be the case for high rigidity chains, the center of mass cutoff is increased to 10.0 Å. Scores based on local atomic density are then assigned to the filtered subset of bonds, promoting the selection of densely packed, interior bonds to approximate bulk-like reactivity. 

\textbf{Solution reactivity:} In different solvation environments, a polymer chain's conformations vary significantly. For example, polymer chains in poor solvent prefer self-interaction as opposed to interaction with the solvent molecules, leading to the polymer chains adopting a tight, globular conformation with chain ends tucked into the center. Atomic density-based scoring metrics are ill-suited for such cases because here, we are interested in the chemical diversity of interfacial reactivity as opposed to bulk-like reactivity. For solvated systems we therefore penalize polymer bonds according to their proximity to chain ends, which encourages sampling from solvent-exposed regions. By ranking bonds based on their maximum distance from the closest chain end, we enrich the dataset with reactions at the polymer-solvent interface, especially for poorly solvated chains.

\subsubsection{Pre-Trimming}
Bulk and solvated chain systems containing more than 800 atoms require a pre-processing step of trimming before the reactive protocol can be applied, in order to minimize the cost of generating AFIR trajectories. At each trimming step, we assure that the reactive bond is re-indexed appropriately. Different trimming protocols are applied to polymers in bulk and solution:

\textbf{Bulk systems:} Bulk chain systems are trimmed down to be at most 500 atoms. At random, a radial cutoff between 4.0 and 6.0 Å is selected to determine the size of spherical, "protected zones" created around each atom in the selected bond for dissociation. Atoms in these zones are off-limits for trimming, so that we can preserve the chemical environment surrounding the dissociation event during data generation. Starting from the chain end furthest from the reactive bond, we iteratively remove one repeat unit at a time, capping the newly exposed atom in the next repeat unit with hydrogen, until either the target atom count is reached or further truncation would encroach upon the protected zones. Substructure matching with RDKit is employed to remove atoms in a given repeat unit at each step. This procedure repeats until the target atom count is achieved, iterating through the chain ends in order of farthest to the reactive bond to closest. If the system iterates through all chain ends and still has the targeted number of atoms, the protected zone cutoff radius is decreased by 1.0 Å, and the process is repeated.

\textbf{Solvated systems:} When more than 150 chain atoms are present, solvated chain systems are trimmed down to at least 800 atoms. This maximum atom threshold is higher than for bulk, so that the reactivity at the polymer-solvent interface may be explored with an adequate number of solvent molecules present. If necessary, solvent molecules are translated to fully wrap the polymer chain in cases where periodic boundaries cause fragmentation. The same trimming algorithm described for bulk systems is then applied, with an additional step where solvent molecules within or overlapping with the protected zones are iteratively kept until atom count thresholds are met. 

\subsubsection{Charge Diversification}
\begin{figure}
    \centering
    \includegraphics[width=0.9\linewidth]{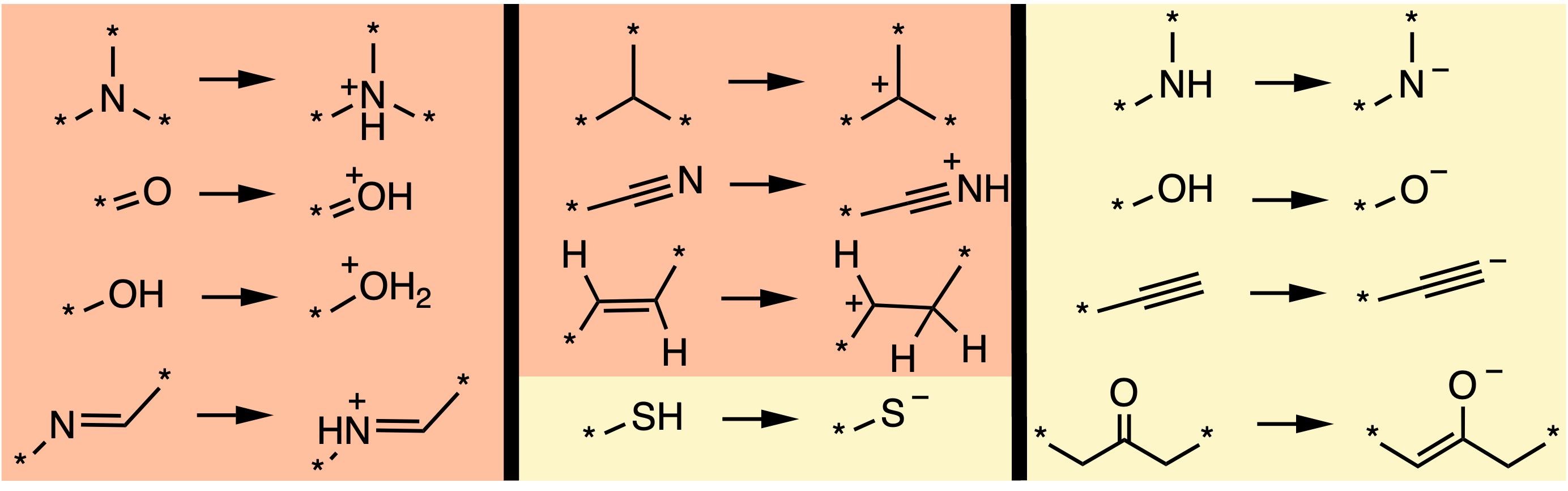}
    \caption{Functional group transformations employed to incorporate net positive (in orange) or negative (in light yellow) charge in a polymer chain prior to AFIR trajectory generation.}
    \label{fig:charged_afir}
\end{figure}

To maximize chemical diversity, one third of all structures undergo protonation while another third undergoes deprotonation to introduce net charge in the system before the AFIR calculation. SMILES arbitrary target specifications (SMARTS) are employed for RDKit substructure matching to identify a subset of functional groups, illustrated in Figure \ref{fig:charged_afir}, present in the system that can undergo these transformations. For each polymer system, all substructure matches corresponding to the available functional groups for a given transformation (protonation or deprotonation) are first enumerated. Only substructure matches that overlap with the reactive region surrounding the bond selected for dissociation (i.e., within 5 Å of the bond atoms) are eligible for selection. A functional group match from the eligible list is then randomly selected to undergo transformation.  The following distinct steps are taken for a given transformation:

\textbf{Protonation:} The position of the new hydrogen atom (H) is computed based on the position and connectivity of the heavy atom to which it will be bonded. A geometric heuristic is applied to place the hydrogen with a bond length of 1.01 Å in a direction according to the heavy atom's number of neighbors ($N$):
\begin{itemize}
    \item $N=3$, H is placed along the normal to the plane defined by the three neighbor atoms, oriented away from the nearest neighbor to avoid poor placement.
    \item $N=2$, H is positioned opposite the bisector of the two neighboring bond vectors. 
    \item $N=1$, H is placed opposite to the existing bonded neighbor.
\end{itemize}
If the generated position of the new hydrogen atom does not fall within 0.5 Å of other atoms, the transformation is applied to the system. If there is overlap, protonation is attempted four more times. After successful protonation, the net charge of the polymer is adjusted accordingly.

\textbf{Deprotonation:} Carbocations are generated through deprotonation in this work, whereas all other deprotonation pathways correspond to deprotonation events. Because hydrocarbons are over-represented in the dataset, the selection probabilities for deprotonation are reweighted. Specifically, when the carbocation-forming SMARTS is present among eligible substructure matches, its selection probability is set to $\frac{1}{14}$. The complementary probability $\frac{13}{14}$ is distributed uniformly among the remaining SMARTS patterns. After a SMARTS is selected according to these modified probabilities, a hydrogen atom within the corresponding substructure match is removed. Bond re-indexing is performed to preserve the identity of the bond selected for dissociation, and the net charge of the polymer system is appropriately modified.

\subsection{AFIR Protocol}
After the above preparation steps, we apply the AFIR method used in OMol25\cite{levine2025openmolecules2025omol25} with a few modifications. We adapt the method to run single-ended searches, in which only the reactant structure and atom indices of the bond breaking are specified. Chain end atoms are fixed in space using ASE constraints. The bond breaking cutoff is increased to 5$\times$ the equilibrium bond length to allow for minimally-guided yet greater exploration along the potential energy surface. A maximum force of 10 eV/Å and a force step of 0.75 eV/Å are used to more efficiently drive reactivity along the polymer chain, compensating for the greater structural inertia of macromolecular-like systems compared to small molecules. The Universal Model for Atoms (UMA)\cite{wood2025uma} trained on the OMol25 dataset is employed to provide energies and forces for AFIR calculations, under the turbo inference setting. AFIR generated frames are then filtered to retain unique structures that capture the reaction pathway, using the algorithm described in OMol25. For solvated chains, the first relaxation with applied force is excluded from structure filtering.

\subsection{Post-Trimming}
To prepare AFIR-generated structures for DFT calculations, each geometry is trimmed to contain less than 250 atoms. The last filtered AFIR structure is used to determine which atoms will be removed from all other structures along the AFIR trajectory, assuming this structure underwent the largest changes in bonding connectivity. Building onto the method used in pre-trimming, protected zones with a radius between 4.0 and 6.0 Å are generated around atoms in the reactive bond as well as all atoms that demonstrate altered connectivity from the starting geometry. As in pre-trimming, atoms are removed one repeat unit at a time– starting with the chain ends fartherest from the reactive site– followed by hydrogen-capping of newly exposed chain ends. Trimming proceeds until either the protected zone is reached or the atom count threshold is satisfied. If the resulting structure still exceeds 250 atoms, the protected zone cutoff is iteratively reduced by 1.0 Å, and the trimming is repeated. The same set of atoms deleted from the last AFIR structure is then removed from all other filtered structures, with appropriate hydrogen capping applied at chain ends where repeat units were extracted.
\section{Extracting Substructures from Trajectories}\label{sec:app:extract}
\subsection{Polymers}\label{sec:app:poly_extract}
From each MD trajectory, we sample 8 frames from the simulated annealing portion and 4 frames from the equilibration portion, resulting in preferential sampling of non-equilibrium polymer structures. Within each of these portions of the MD trajectory, we calculate the root mean square displacement (RMSD) between all polymer structures and sample the frames of the trajectory with the largest dissimilarity to all other structures in the trajectory. In the ML MD and DFTB simulations, frames are randomly sampled from across the trajectory. After sampling a frame from a trajectory, a polymer residue is selected at random and either
a spherical shell or a "cylindrical" shell (including several neighboring residues and their close contacts) is selected up to a randomly chosen maximum number of atoms (150, 200, or 300). These shells include entire monomers and therefore terminate at the connection point between monomer repeat units. This shell is extracted and capped with hydrogen atoms. We ensure that the extracted substructure is appropriately contracted across periodic boundaries to a compact form suitable for molecular DFT

\subsection{Lipids}\label{sec:app:lipids_extract}
For lipid systems, we cycled through head, linker, and tail moieties of randomly selected lipid molecules in an MD simulation, attempting to extract clusters of them and their environment. Each moiety’s environment was defined as all other heads, linkers, or tails with any atom within 3 Å of the center moiety but excluding any atoms more than 10, 5, or 4 Å from the center moiety (opting for the largest such cluster within the maximum atom limit of 300 atoms). Cut bonds were capped with H’s except for phosphate groups which were capped with -OMe groups. Each of the 674 MD simulations was uniformly sampled every 10 ns and 30 structures were obtained by this procedure from each simulation snapshot.

\section{Additional Analyses}\label{sec:app:add_analyses}

\subsection{OPoly26 Test Set Force Errors}\label{sec:test_force_errors}

\newcolumntype{Y}{>{\centering\arraybackslash}X}
\begin{table}[htbp]
\centering
\caption{Breakdown of force prediction errors on the OPoly26 composition test split. For each model, we report the mean absolute error of forces (meV/\AA) on various polymer types, as defined in Figure \ref{fig:omers_summary}. The number below each polymer type indicates the number of DFT snapshots corresponding to that polymer class in the test set.}
\label{tab:test_force_errors}
\scriptsize 
\setlength{\tabcolsep}{2pt} % Slightly tighter for very wide tables
\begin{tabularx}{\textwidth}{@{} l *{8}{Y} @{}}
\textbf{Model} 
& \makecell[b]{\textbf{High Entropy} \\ \textbf{Copolymers} \\ (N=115,866)}
& \makecell[b]{\textbf{Alt.} \\ \textbf{Copolymers} \\ (N=36,592)}
& \makecell[b]{\textbf{Lipids} \\ (N=8,676)}
& \makecell[b]{\textbf{Peptoids} \\ (N=47,595)}
& \makecell[b]{\textbf{Reactivity} \\ (N=18,462)}
& \makecell[b]{\textbf{Homopolymers} \\ (N=5,702)}
& \makecell[b]{\textbf{Solvated} \\ \textbf{Polymers} \\ (N=8,878)}
& \makecell[b]{\textbf{Ion-Inserted} \\ \textbf{Polymers} \\ (N=6,620)} \\
\midrule
OMol25 Only
& 4.4 & 4.2  & 3.8  & 4.4  & 46.3  & 3.6  & 3.2  & 6.5  \\
\midrule
UMA-s-1p1 
& 4.9 & 5.0  & 4.4  & 4.8  & 49.0  & 4.2  & 3.7  & 7.2  \\
\midrule[2pt]
OPoly26 Only 
& 2.8 & 2.6  & 2.7  & 2.6  & 21.3  & 2.4  & 2.2  & 5.6  \\
\midrule
OPoly26 + OMol25 
& 3.5 & 3.3  & 3.0  & 3.5  & 22.2  & 3.0  & 2.7  & 5.5  \\
\midrule
UMA-s-1p2 
& 3.0 & 2.8  & 2.4  & 2.9  & 20.6  & 2.6  & 2.3  & 4.5  \\
\bottomrule
\end{tabularx}
\begin{tablenotes}
  \item \kindatiny Force MAE(meV/\angs)
\end{tablenotes}
\end{table}

\end{document}